\documentclass{ptephy}

\usepackage{amsmath} 
\usepackage{graphics} 

\usepackage{graphicx}
\usepackage{amssymb}
\makeatletter

 \@addtoreset{equation}{section}
\makeatother
\newcommand{\beq}{\begin{eqnarray}}
\newcommand{\eeq}{\end{eqnarray}}
\usepackage{color}

\begin{document}

\title{A novel scheme for the wave function renormalization of the composite operators}

\author{\name{Etsuko Itou}{1}}

\address{\affil{1}{High Energy Accelerator Research Organization (KEK), Tsukuba 305-0801, Japan}
\email{eitou@post.kek.jp}}

\begin{abstract}%
We propose a novel renormalization scheme for the hadronic operators.
The renormalization factor of the operator in this scheme is normalized by the correlation function at tree level in coordinate space.
If we focus on the pseudo scalar operator, then its renormalization factor is related to the mass renormalization factor of the fermion through the partially conserved axial-vector current (PCAC) relation.
Using the renormalization factor for the pseudo scalar operator in our scheme, we obtain the mass anomalous dimension of the SU($3$) gauge theory coupled to $N_f=12$ massless fundamental fermions,
which has an infrared fixed point (IRFP).
The mass anomalous dimension at the IRFP is estimated as 
$\gamma_m^*= 0.044 \hspace{3pt}_{-0.024}^{+0.025} (\mbox{stat.}) \hspace{3pt} _{-0.032}^{+0.057} (\mbox{syst.})$.
\end{abstract}

\subjectindex{B32, B38, B44}

\maketitle

\section{Introduction}
Lattice gauge theory provides a regularization method for the gauge theory.
To regulate the theory, we introduce a lattice spacing ($a$) as a ultraviolet (UV) cutoff and a finite lattice extent as an infrared (IR) cutoff.
For the lattice gauge theory, there are several useful renormalization schemes for the gauge coupling constant, {\it e.g.} the Schr\"{o}dinger functional (SF) scheme~\cite{Luscher:1991wu}, the potential scheme~\cite{Campostrini:1995aa}, the Wilson loop scheme~\cite{Bilgici:2009kh},
the twisted Polyakov loop (TPL) scheme~\cite{deDivitiis:1993hj,TPL}, the Wilson flow (Yang-Mills gradient flow) scheme~\cite{Fodor:2012td, Fritzsch:2013je} etc.
A variety of renormalization scheme for composite operators has also been given,  {\it e.g.} the SF scheme~\cite{Capitani:1998mq} and RI-MOM scheme~\cite{Martinelli:1994ty} and so on.
Concerning the fermion mass renormalization, the quark mass renormalization factor is related to that of the pseudo scalar operator because of the partially conserved axial-vector current (PCAC) relation.

In this paper, we propose a novel scheme for the composite operators.
The basic idea is to normalize the renormalization factor using the tree level correlation function of the operator in coordinate space.
A similar renormalization scheme is provided in the paper~\cite{Cichy:2012is}. 
In this paper, we give a scheme with the twisted boundary condition and give an explicit value of the tree level correlator for the pseudo scalar operator on the lattice.
Thanks to the twisted boundary condition, we can obtain it at the massless point.

We also apply this renormalization scheme to derive the anomalous dimension of the pseudo scalar operator for the SU($3$) gauge theory coupled to $N_f=12$ massless fermions.
In our previous work~\cite{TPL}, we investigated the running coupling constant of this theory using the twisted Polyakov loop scheme from the perturbative region to the IR region.
We found the growth of the renormalized coupling halts in the IR region, which verifies that the infrared fixed point (IRFP) exists in this theory.
At the IRFP, we expect that an interactive conformal field theory is realized.
Note that the lattice gauge action is defined at the Gaussian UV fixed point and we do not know the explicit form of the action of such interactive IR conformal theory.
However, we expect that  the theory is sufficiently close to the IRFP in the region, where the coupling constant does not show the growth when the energy scale changes.
Here we declare that the theory on the lattice realizes the conformal fixed point theory.

Conformal fixed points are the most important object in the quantum field theories.
At the conformal fixed point the critical exponents ({\it e.g.} the anomalous dimension of the operators) are the scheme independent quantities, and
these exponents classify the universality class of the quantum field theories.
Among the several critical exponents, the one that is related to the relevant operator is crucial to define the IR field theory. 
In this paper, we determine numerically the universal mass anomalous dimension of the interactive conformal field theory.

Recently, several methods to obtain the mass anomalous dimension for the conformal gauge theory realized at the IRFP in many-flavor SU($N_c$) gauge theories have been proposed.
The step scaling is one of the methods based on the renormalization group for the finite scaling, and this method can be applied to the non-conformal field theories~\cite{Luscher:1991wu}.
The other method is to use the hyperscaling~\cite{Miransky:1998dh} for the mass deformed conformal gauge theory.
The application of the hyperscaling on the lattice was pointed out by M.~A.~Luty and F.~Sannino~\cite{Luty:2008vs, Sannino:2008pz}, and the detailed practical discussions of mass deformed conformal gauge theory have been shown by L.~Del Debbio and R.~Zwicky~\cite{DelDebbio:2010jy}.
This method is based on the assumption of the existence of the interactive conformal field theory, where the scaling of the operators is different from the Gaussian (canonical) case.
The mass anomalous dimension is derived by  the fit of the mass spectrum of hadronic state or the chiral condensate in a small mass region.
A similar method to estimate the mass anomalous dimension using the fit for the massless SU($3$) gauge theory has also been proposed~\cite{deForcrand:2012vh,Cheng:2013eu}.
They utilized the massless fermion, and from the scaling of ($1/L$), where $L$ is a finite lattice extent, they estimated the mass anomalous dimension. 
The independent method has been suggested in the paper~\cite{Ishikawa:2013wf}. They assumed that the correlation function in the finite volume around the IRFP became the Yukawa-type function and derived the anomalous dimension from the fit.

In this work, we obtain the mass anomalous dimension at the IRFP using the step scaling method.
In our numerical simulation, we introduce the twisted boundary condition for both gauge field and the fermion field.
The boundary condition kills the zero mode contribution and regularizes the fermion matrix even in the massless case.
Thus we carry out the simulation using exactly massless fermions.
Several independent groups have been investigating the mass anomalous dimension of the SU($3$) gauge theory coupled to $N_f=12$ fermions~\cite{Cheng:2013eu, Appelquist:2011dp, DeGrand:2011cu,Aoki:2012eq}.
The works~\cite{Appelquist:2011dp} -- \cite{Aoki:2012eq} are based on the hyperscaling method for mass deformed conformal gauge theory applied to the simulation with massive fermions.
In the paper~\cite{Cheng:2013eu}, they utilize the (approximately) massless fermion, and derive the universal mass anomalous dimension in the infinite volume limit using the hyperscaling for the Dirac eigenmodes. 
This work is the first study on the mass anomalous dimension for the SU($3$) $N_f=12$ massless gauge theory using the step scaling method.
We expect that the value of the mass anomalous dimension at the IRFP is independent of the derivation.

This paper is organized as follows:
In Sec.~\ref{sec:definition}, we give the definition of a novel renormalization scheme for the composite operators.
The renormalization factor of the pseudo scalar operator is related to the fermion mass renormalization factor, therefore in the rest of the paper we focus on the pseudo scalar operator and determine the fermion mass anomalous dimension.
In Sec.~\ref{sec:step-scaling}, we show the strategy to obtain the mass anomalous dimension using the step scaling method and give a definition of the step scaling function in our scheme.
Note that there are two definitions of the step scaling function in our scheme, since there are two scales in the observable for the renormalization condition.
In Sec.~\ref{sec:section-simulation-detail}, we explain our numerical simulation setup.
We compute the correlation function at tree level, which is needed to define the renormalization factor in our new scheme in Sec.~\ref{sec:tree-level} and Appendix~\ref{sec:shape-correlation}.
We determine the mass anomalous dimension at the IRFP of the SU(3) gauge theory coupled to $N_f=12$ massless fermions in Sec.~\ref{sec:anomalous-dimension}.
We find that the mass anomalous dimension at the IRFP is given by
\beq
\gamma_m^*&=& 0.044 \hspace{5pt}_{-0.024}^{+0.025} (\mbox{stat.}) _{-0.032}^{+0.057} (\mbox{syst.}), 
\eeq
where the systematic error includes the uncertainties coming from the continuum extrapolation and the value of the coupling constant at the IRFP.
We discuss the comparison with the other works in Sec.~\ref{sec:discussion}.
We conduct a discussion about the promising methods to studying such universal quantity around the IRFP using the lattice simulation in Sec.~\ref{sec:summary}.

\section{A novel renormalization scheme for the anomalous dimension}\label{sec:definition}
We give a new renormalization scheme of an arbitrary composite operator ($H$)\footnote{A similar scheme with a different boundary condition is independently provided in paper~\cite{Cichy:2012is}}.
In renormalizable theories, a nonperturbative renormalized coupling constant can be defined by amplitudes of the observables.
The SU($3$) gauge theory coupled to a small number ($N_f \le 16$) of fundamental fermions is asymptotically free and it is described by two kinds of parameter: the gauge coupling and the mass parameter of the fermions.

The renormalization factor can be defined by the correlator of the bare operator ($H$),
\beq
C_{H}(t)&=& \sum_{\vec{x}} \langle H(t,\vec{x}) H(0,\vec{0}) \rangle.
\eeq
To obtain the finite renormalized value of the correlator, we introduce a nonperturbative renormalization factor ($Z_H$) as follow:
\beq
C_{H}^R (t) =Z_H^2 C_{H} (t).
\eeq
Here $C_{H}^R$ denotes a renormalized correlation function and it is finite.
On the other hand, the renormalization factor $Z_H$ and the nonperturbative bare correlation function diverge, and on the right hand side these divergences are canceled each other.

We introduce the renormalization condition on the renormalized correlator, in which the renormalized correlator is equal to the tree level amplitude:
\beq
C_{H}^R(t)=C_{H}^{\mathrm{tree}} (t).\label{eq:rg-condition}
\eeq
The renormaliation factor of the composite operator is thus defined by
\beq
Z_H= \sqrt{ \frac{C_H^{\mathrm{tree}}(t)}{C_H (t)} },
\eeq
at the fixed propagation length ($t$) in coordinate space.
Thus the factor $Z_H$ is normalized by the tree level value for each propagation length.

On the lattice, the nonperturbative bare correlation function ($C_{H} (t)$) is calculated by lattice numerical simulations.
The correlation function on the lattice depends on the propagation time ($t/a$), the bare coupling constant ($g_0$), the bare mass ($m_0$) and the lattice size($L/a,T/a$).
Thus it is denoted by $C_H(g_0, m_0a,t/a,T/a,L/a)$ on the lattice.
We fix the ratio between the temporal and the spatial lattice extents, and identify an inverse of lattice spatial extent ($1/L$) as a renormalization scale ($\mu$).
Then, we give the definition of $Z$ factor in this scheme on the lattice, 
\beq
Z_H(g_0, m_0a ,t/a, a/L) =\sqrt{\frac{C_{H}^{\mathrm{tree}} (g_0, m_0a, t/a, a/L)}{C_{H} (g_0, m_0a, t/a, a/L)}}. \label{eq:def-Z}
\eeq

Let us define these nonperturbative renormalized parameters at the energy scale ($\mu$) as $\bar{g}^2 (\mu) \equiv Z_g g_0^2$ and $\bar{m} (\mu) \equiv Z_m m_0$.
Here  $Z_g$ and $Z_m$ denote nonperturbative renormalization factors for each parameter.
These factors are functions of dimensionless parameters $g_0$,$m_0a$ and $a\mu$, so that they can be written by $Z_g=Z_g (g_0,m_0a, a\mu)$ and $Z_m=Z_m (g_0, m_0a, a\mu)$
\footnote{Here we simply consider a multiplicative renormalization factor of the mass. Actually, in our simulation we use the staggered fermion, so that there is no additive mass. Of course one can use this renormalization scheme in existence of the additive mass renormalization. }.
In QCD, the renormalized mass is also defined through the partially conserved axial-vector current (PCAC) relation:
\beq
\partial_\mu (A_R)_\mu =2 \bar{m} P_R,
\eeq
where $A_R$ and $P_R$ denote the renormalized axial-vector current ($A_{R\mu}(x)=Z_A \bar{\psi}(x) \gamma_\mu \gamma_5 \psi (x)$) and pseudo scalar operator ($P_R(x)=Z_P\bar{\psi}(x)\gamma_5 \psi$) respectively.
Here we introduce the renormalization factors for these operators.
Note that $Z_A(g_0,m_0)$ is scale independent because the axial current is renormalized through current algebra.
Thus the PCAC relation gives the relationship between the mass renormalization factor ($Z_m$) and the renormalization  factor ($Z_P$) of the pseudo scalar operator as,
\beq
Z_m (\mu) = Z_A Z_P^{-1} (\mu),\label{eq:Zm-Zp}
\eeq
at each renormalization scale.
The anomalous dimension of the dimensionless running mass of the fermions, 
\beq
\mu d \bar{m}/d \mu = - \gamma_m(\bar{g},\bar{m}) \bar{m},
\eeq
can be calculated from the scale dependence of the pseudo scalar renormalization factor $Z_P$ as
\beq
\gamma_m = \frac{d \ln Z_P}{d \ln \mu}.\label{eq:def-gamma-m}
\eeq
In this paper, we study on the massless fermion theory, and measure the renormalization factor of the pseudo scalar operator.

The definition is basically applicable to any boundary conditions, and the one for the periodic boundary is provided in the paper~\cite{Cichy:2012is}. 
In this paper we will introduce the twisted boundary condition in $x$ and $y$ directions, and it allows us to study the exact massless case.

\section{Step scaling function}\label{sec:step-scaling}
Let us consider the scale dependence of the renormalization factor $Z_P$ (Eq.~(\ref{eq:def-gamma-m})).
Here we assume the massless renormalized fermion.
First, we introduce the discrete mass step scaling function from the factor $Z_P$.

In our renormalization scheme, the renormalization factor $Z_P$ has two independent scales, the propagation time ($t$) and lattice temporal size ($T$)
\footnote{We always fix the ratio of $T$ and $L$.}.
To see the scale dependence of the factor $Z_P$, there are two definitions of the scaling function.
One definition of the scaling function is given by the factor $Z_P$ at a fixed ratio $r=t/T$, which we call as ``fixed $r$" definition. 
Here $r$ takes a value $0< r \le1/2$ because of the periodic boundary condition on the lattice. 
We obtain the scale dependence of the factor $Z_P$ when we change both the physical propagation length and lattice size together.
The other definition is given by the factor $Z_P$ at a fixed $t$.
We change the dimensionless quantity $r$ and the temporal physical lattice extent $T$ in the latter definition.
In the former definition, the renormalization scale is parametrized by the lattice spatial size.
In the latter case, if $T \gg L$ and $t > L$, then the renormalization scale is given by $1/t$.
In this paper, we use the former ``fixed $r$" definition.
Now, the factor $Z_P$ in Eq.~(\ref{eq:def-Z}) depends only on the bare coupling constant and the lattice spatial size: $Z_P(g_0,a/L)$ in a fixed $r$ scheme.
From now on we use $\beta$ to denote the bare coupling constant and $\beta=6/g_0^2$.

We give a comment on a choice of the parameter $r$ in ``fixed $r$" scheme.
If we choose small $r$, the lattice data might suffer from a large discretization effect.
On the other hand, in the case of large $r$, the signal of correlators of the hadronic operators might become noisy except for the lightest state.
Practically, we have to search for the optimal range of $r$.
Note that at the fixed point, the anomalous dimension is a scheme independent quantity, so that it should be independent of $r$.
We discuss $r$ dependence of our result in Sec.~\ref{sec:r-dependence}.

\begin{figure}[h]
\begin{center}
   \includegraphics[width=6cm,clip]{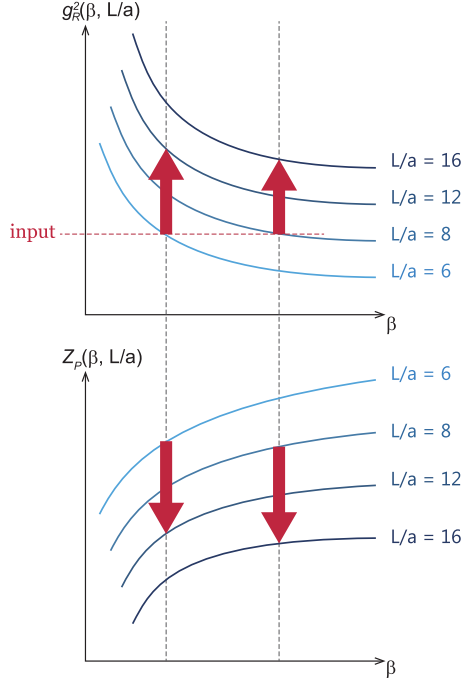}
  \caption{The strategies of the step scaling method for the coupling constant and the renormalization factor for the operator $P$. We measure the growth ratio of each quantity at fixed $\beta$.}
  \label{fig:strategy}
\end{center}
\end{figure}

Now, we give a brief review of the strategy to obtain the mass anomalous dimension using the step scaling method.
The idea is established by ALPHA collaboration~\cite{Capitani:1998mq}.
Figure~\ref{fig:strategy} shows a schematic picture which describes the strategy of step scaling for the renormalized coupling constant and the renormalization factor of operators.
The top panel shows the step scaling for the renormalized coupling.
To obtain the scale dependence of the renormalized coupling in a renormalization scheme, we measure the growth ratio of renormalized coupling when the lattice extent becomes $s$ times with fixed value of bare coupling constant.
Practically we carry out the following procedures.
First, we choose a value of renormalized coupling constant $u=g_R^2 (1/L)$ and tune the value of $\beta$ to realize $u$ for each lattice size.
Next, we measure the renormalized coupling constant with the tuned value of $\beta$ at the larger lattice $sL/a$.
The renormalized coupling constant on the larger lattice is called the discrete step scaling function: $\Sigma(u)$. Here $s$ is the step scaling parameter ($1 < s$).
Finally we take a continuum limit of the discrete step scaling function: $\sigma(u)=\lim_{a \rightarrow 0} \Sigma(u)=g^2_R(\mu=1/sL)$.
The growth ratio of the renormalized coupling ($\sigma(u)/u$) essentially gives a discrete beta function.

To obtain the scale dependence of the renormalization factor of a operator, we measure the growth ratio of the factor $Z_P$. 
If the operator is a pseudo scalar operator, it is called the discrete mass step scaling function because of the relationship Eq.(\ref{eq:Zm-Zp}).
The explicit definition of the mass step scaling function is given by
\beq
\Sigma_P (\beta , a/L; s)&=&  \frac{Z_P (\beta, a/sL)}{Z_P(\beta, a/L)} \mbox{ at $m(\beta)=0$.}\label{eq:disc-sigma}
\eeq
The mass step scaling function on the lattice includes the discretization error.
To remove it, we take the continuum limit ($a \rightarrow 0$) keeping the renormalized coupling ($u=g_R^2 (1/L)$) constant.
\beq
\sigma_P (u,s) &=& \left. \lim_{a \rightarrow 0} \Sigma_P (\beta, a/L;s) \right|_{u=const}.\label{eq:def-sigma}
\eeq

In the continuum limit, this mass step scaling function is related to the mass anomalous dimension as,
\beq
\sigma_P (u,s) =\left( \frac{u}{\sigma(u)} \right)^{d_0/2b_0} \exp \left[ \int_{\sqrt{u}}^{\sqrt{\sigma(u)}} dx \left( \frac{\gamma_m(x)}{\beta(x)} + \frac{d_0}{b_0 x} \right) \right],
\eeq
where $\gamma_m(x)$ and $\beta(x)$ denote the mass anomalous dimension and the beta function respectively, and $d_0$ and $b_0$ denote coefficients of them in $1$-loop order  as follows:
\beq
\beta(x)&=&- x^3 \left[ b_0 +b_1 x^2 +b_2 x^4 +O(x^6)  \right], \nonumber\\
\gamma_m(x)&=& x^2 \left[ d_0 +d_1 x^2 +O(x^4) \right].
\eeq 

This relation becomes simple when the theory is conformal:
\beq
\int_{\bar{m}(\mu)}^{\bar{m}(\mu/s)} \frac{dm}{m} = -\gamma_m^* \int_{\mu}^{\mu/s}\frac{dq}{q},
\eeq
and we can estimate the anomalous dimension at the fixed point with the following equation:
\beq
\gamma_m^*(u^*)=-\frac{\log |\sigma_P (u^*,s)|}{\log|s|}.
\eeq
Here $u^*$ denotes the fixed point coupling constant.

Note that in Eq.~(\ref{eq:def-sigma}), there is a freedom of the choice of the renormalization scheme concerning the input renormalized coupling constant. 
We use the set of the bare coupling constant and the lattice size to realize the input renormalized coupling constant ($u$) with a renormalization scheme for the gauge coupling constant.
The energy scale is defined by the input renormalized coupling, and the energy dependence of the mass step scaling function comes through the renormalized coupling constant.
We can use any combinations of the renormalization schemes for the renormalized coupling and the wave function renormalization.
Generally, the value of $\sigma_P (u)$ and the mass anomalous dimension depends on the choice of the renormalization schemes.
At the fixed point, although the value of $u^*$ depends on the renormalization scheme, the mass anomalous dimension is independent of the renormalization schemes of both the mass and the coupling constant.

\section{Simulation setup}\label{sec:section-simulation-detail}
The gauge configurations are generated by the Hybrid Monte Carlo algorithm, and we use the Wilson gauge and the naive staggered fermion actions.
We introduce the twisted boundary conditions for $x, y$ directions and impose the usual periodic boundary condition for $z, t$ directions, which is the same setup with our previous work~\cite{TPL}.

For the link valuables ($U_\mu$), we introduce the following twisted boundary condition in $x$ and $y$ directions  on the lattice:
\beq
U_{\mu}(x+\hat{\nu}L/a)=\Omega_{\nu} U_{\mu}(x) \Omega^{\dag}_{\nu},  \label{twisted-bc-gauge}
\eeq
for $\mu=x,y,z,t$ and $\nu=x,y$.
Here, $\Omega_{\nu}$ ($\nu=x,y$) are the twist matrices which have the following properties:
\beq
 \Omega_{\nu} \Omega_{\nu}^{\dag}=\mathbb{I},
 (\Omega_{\nu})^3=\mathbb{I},
 \mbox{Tr}[\Omega_{\nu}]=0, \nonumber
\eeq
and 
\beq
\Omega_{\mu}\Omega_{\nu}=e^{i2\pi/3}\Omega_{\nu}\Omega_{\mu},
\eeq
 for a given $\mu$ and $\nu$($\ne \mu$).

For the fermions, we identify the fermion field as a $N_c \times N_s$ matrix ($\psi^a_\alpha$($x$)), where $a$ ($a=1,\cdots,N_c$) and $\alpha$ ($\alpha=1,\cdots,N_s$) denote the indices of the color and smell. 
We can then impose the twisted boundary condition for fermion fields as
\beq
\psi^a_{\alpha} (x+\hat{\nu}L/a)= e^{i \pi/3} \Omega_{\nu}^{ab} \psi^{b}_{\beta} (\Omega_{\nu})^\dag_{\beta \alpha}\label{eq:fermion-bc}
\eeq
for $\nu=x,y$ directions.
Here, the smell index can be considered as a part of  ``flavor'' index, so that the number of flavors should be a multiple of $N_s$, in our case $N_s$ should be the multiple of $N_c=3$.

Because of the twisted boundary conditions the fermion determinant is regularized even in the massless case, so that we carry out an exact massless simulation to generate these configurations.
The simulation are carried out with several lattice sizes ($L/a=6,8,10,12,16$ and $20$)
at the fixed point of the renormalized gauge coupling in the TPL scheme~\cite{TPL}.
In this simulation, we fix the ratio of temporal and spatial directions: $T/a=2L/a$
\footnote{The new renormalization scheme can be defined using $T/a=L/a$ lattice. However, our simulation uses the staggered fermion which takes a value only on even-number space-time sites. That decrease the number of data points. Actually we carried out the simulation and the step scaling using $T/a=L/a$ lattices, but we found there is a quite large scaling violation. That is the reason why we use the extended ($T/a=2L/a$) lattices in this work.}.
We use the tuned value of $\beta$ where the TPL coupling is the fixed point value for each $(L/a)^4$ lattices.
We neglect the possibility of induced scale violation coming from the change of lattice volume $(L/a)^4 \rightarrow 2(L/a)^4$ since the renormalized coupling constant on $(L/a)^4$ is the same with one on $(2L/a)^4$ after taking the continuum limit.
We generate $30,000$--$80,000$ trajectories for each $(\beta, L/a)$ combination, and measure the pseudo scalar correlator with every $100$ trajectories.
The statistical error is estimated using bootstrap method.
Thus we obtain the data of $Z_P$ for single configuration, and randomly resample the set of data $O(1,000)$ times for each lattice parameter.

Now, we explain the detailed values of $\beta$ in our simulation.
In this paper, we focus on the mass anomalous dimension at the IRFP.
In our previous paper~\cite{TPL}, we found that the existence of the IRFP at 
\beq
g_{\mathrm{TPL}}^{*2} = 2.686 \pm 0.137 (\mbox{stat.}) ^{+0}_{-0.160} (\mbox{syst.}),
\eeq
in the TPL scheme.
The parameter sets which realize the fixed point coupling on the lattice is shown in Table.~\ref{table:beta-L}.
\begin{table}[h]
\begin{center}
\begin{tabular}{|c|c|c|c|c|}
\hline
{} & {} &$g_{\mathrm{TPL}}^{2}=2.475$  & $g_{\mathrm{TPL}}^{2}=2.686$ & $g_{\mathrm{TPL}}^{2}=2.823$ \\     
\hline \hline
L/a & T/a & $\beta$ & $\beta$ & $\beta$  \\  
\hline
6   &  12 & 5.378          &  4.913           & 4.600         \\
8   &  16 & 5.796          &  5.414           & 5.181              \\
10   &  20 &  5.998          &  5.653           & 5.450            \\
12   &  24 &  6.121          &  5.786           & 5.588              \\
16   &  32 &  6.241          &  5.909           & 5.709            \\
20   &  40 &6.296          &  5.944           & 5.723             \\
\hline
\end{tabular}
\caption{ The values of $\beta$ for each $L/a$ which give the TPL coupling constant at the IRFP.} \label{table:beta-L}
\end{center}
\end{table}
We generate the configurations using these parameters on $(L/a)^3 \times 2L/a$ lattices.
Furthermore, to carry out the step scaling we also add the parameter sets which is shown in Table.~\ref{table:beta-L2}.
\begin{table}[h]
\begin{center}
\begin{tabular}{|c|c|c|c|c|}
\hline
L/a & T/a & $\beta$ & $\beta$ & $\beta$  \\     
\hline 
8   & 16 &  5.378          &  4.913           & 4.600         \\
10   &  20 & 5.378          &  4.913           & 4.600         \\
12   &  24 & 5.378          &  4.913           & 4.600         \\
16   &  32 & 5.796          &  5.414           & 5.181            \\
20   &  40 & 5.998          &  5.653           & 5.450             \\
\hline
\end{tabular}
\caption{ Additional simulation parameters.} \label{table:beta-L2}
\end{center}
\end{table}

The pseudo scalar correlator can be presented by the fermion propagators ($S(t,\vec{x})$)
\beq
C_{PS}(t)&=&\langle \sum_{\vec{x}} \bar{\psi}(t,\vec{x}) \gamma_5 \otimes \gamma_5 \psi(t,\vec{x}) \bar{\psi}(0,\vec{0}) \gamma_5 \otimes \gamma_5 \psi(0,\vec{0})  \rangle,\\
&=&\sum_{\vec{x}} \langle tr \left[ S(t,\vec{x}) S^{\dag}(t,\vec{x}) \right] \rangle, \label{eq:def-c-pi}
\eeq
where $\gamma_5 \otimes \gamma_5$ specifies the spin and flavor structure.
We measure the pseudo scalar correlator using the point source at $t=0$. 
We construct the Dirac field by the staggered fermion($\chi$) on the hypercubic space-time.
The pseudo scalar correlator calculated by the two point function of staggered fermion ($C_\chi(t) \equiv \sum_{\vec{x}} |\langle  \bar{\chi}(t,\vec{x}) \chi (0,\vec{0} )  \rangle |^2$) has the value at even temporal sites:
\beq
C_{PS}(2t) = 32 \left[ 2C_{\chi}(2t) +C_{\chi}(2t+1) +C_{\chi}(2t-1) \right].
\eeq
Practically, we introduce a tiny bare fermion mass $ma=10^{-5}$--$10^{-6}$ for the measurement of the correlators.
To check the smallness of this bare quark mass rather than the twisted momentum even in the strong coupling region, we have changed the mass to $ma=10^{-7}$ and confirmed that the effect of the mass is negligible.

\section{Calculus of the correlator at tree level with twisted boundary condition} \label{sec:tree-level}
We compute the correlator at the tree level to normalize the renormalization factor in the renormalization condition Eq.~(\ref{eq:rg-condition}).
The correlation function of the pseudo scalar at the tree level corresponds to the correlation function of the two free fermions.
We can calculate it using the vacuum configurations.
There are three possible choices of the vacuum configurations in the case of SU($3$) gauge theory, since the pure SU($N_c$) gauge theory has $Z_{N_c}$ global symmetric degenerated classical vacua at $U_{\mu}=\exp(2\pi i \theta_\mu/N_c)$, where $\theta_\mu=0,1, \cdots, N_c-1$ for each direction.
According to the semi-classical analysis of the $1-$ loop effective potential (See Sec.~4 in Ref.~\cite{TPL}), 
the $3^4-$fold degenerated vacua, where the Polyakov loop in $z,t$ directions has a nontrivial complex phase ($\exp(\pm 2 \pi i/3)$) are chosen in our lattice setup.
Therefore we use that the vacuum configuraions, $U_\mu = \exp(\pm 2 \pi i /3\hat{T}) {\mathbb I}$ for $\mu=z,t$, to derive the correlator at the tree level.
For $x$ and $y$ directions, the effective potential does not depend on the choice of the $\theta_\mu$. 
We use the simple constant configuration $U_{\mu}=\mathbb{I}$.

\begin{figure}[h]
\begin{center}
   \includegraphics[width=8cm]{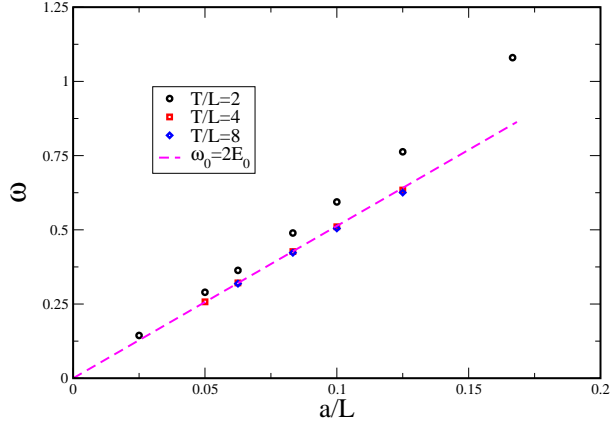}
\vspace{1cm}
 \caption{The value of $\omega$ when we solve the function $C(\hat{t})=c_0 \cosh (\omega (\hat{t}-\hat{T}/2))$ using two independent data around $\hat{t}=\hat{T}/2$. The dot line denotes $\omega_0 = 2E_0$, where $E_0$ denotes the lowest energy of single free fermion given by Eq.(\ref{eq:tree-E0}).}
 \label{fig:omega0}
\end{center}
\end{figure}
The data of the pseudo scalar correlator for each lattice size using this vacuum configuration is shown in Table~\ref{table:tree-Ct} in Appendix~\ref{sec:app-raw-data}.
The data can be fitted by $\cosh[\omega(\hat{t}-\hat{T}/2)]$ well in long $\hat{t}$ region (See  Appendix~\ref{sec:shape-correlation}), where the hatted symbol denotes the quantities in the lattice units.
Let us consider the meaning of ``$\omega$" in the massless fermion case.
Figure~\ref{fig:omega0} shows the values of the fitted parameter ``$\omega$" for several lattice extents $(L/a)^3 \times (T/a)$.
Here we measure the tree level correlator in the cases of $T/L=2,4,8$ for $L/a=6,8,12,16,20$ and $40$.
For each $T/L$, the parameter ``$\omega$", which we call the effective mass, is proportional to $a/L$.

At the long distance, only the lowest energy mode must survive, and we expect that the correlation function can be approximated as $C(t) \sim e^{-2 E_0 t}$, where $E_0$ is the lowest energy of single fermion.
It is realized at the lowest energy state those four-dimensional momentum is zero: $p^2=E_0^2 -\vec{p}^2=0$.
Thus the lowest energy is obtained by the sum of lowest spatial discrete momenta.
In our lattice setup the momentum for each direction is given in Eq.(22) in the paper\cite{TPL}:
\beq
\hat{p}_\mu &=& \frac{2\pi}{\hat{L}} n_\mu +\frac{\pi (2m_\mu+1) }{3 \hat{L}} \mbox{ \hspace{2pt}  for $\mu=x,y$},\\
\hat{p}_\mu &=& \frac{2\pi}{\hat{L}} n_\mu +\frac{ 2 \pi}{3 \hat{L}} \mbox{ \hspace{50pt} for $\mu=z$},
\eeq
where $n_\mu=0, 1, \cdots, \hat{L}/2-1$ and $m_\mu=0,1,\cdots, N_c-1$ with $(m_x^\perp, m_y^\perp) \ne (0,0)$.
The lowest energy of single fermion is analytically calculated as the following,
\beq
E_0^2= \sum_{i=1}^3 \vec{p}_i^2 &=&  \left(\frac{\pi}{3\hat{L}} \right)^2 + \left(\frac{\pi}{3\hat{L}} \right)^2 + \left(\frac{2\pi}{3\hat{L}} \right)^2  \nonumber\\
&=& \left( \frac{\sqrt{6} \pi}{ 3 \hat{L}} \right)^2 .\label{eq:tree-E0}
\eeq
The dot line in Fig.~\ref{fig:omega0} denotes the line of $\omega_0 =2E_0$.
We find that around $t/a=T/2a$ with $T/L \ge 4$ only the lowest mode remains.
In this paper, we uese $T/L=2$ lattices, so that there are still some contributions from the second lowest energy and the higher modes in the correlator at the tree level.
However, the lattice data in Fig.~\ref{fig:omega0} shows that the effective mass $\omega$ vanishes in the continuum limit even if such modes remains.

\section{Mass anomalous dimension at the IRFP}\label{sec:anomalous-dimension}
We measure the mass anomalous dimension at the IRFP for the SU($3$) $N_f=12$ gauge theory using the step scaling method.
When we take the continuum limit of the mass step scaling function, we use the TPL coupling constant at the IRFP as an input renormalized coupling.
In the paper~\cite{TPL}, the value of TPL coupling at the IRFP is determined
\beq
g_{\mathrm{TPL}}^{*2} =  2.686 \pm 0.137 (\mbox{stat.}) ^{+0}_{-0.160} (\mbox{syst.}).\label{eq:TPL-g-value}
\eeq
We take the mass anomalous dimension at the central value, ($g_{\mathrm{TPL}}^{*2}=2.686$) as a central analysis using the $s=1.5$ step scaling for the ``fixed $r$" scheme with $r=1/2$.
We estimate the systematic error by taking the discrepancy among several kinds of the continuum extrapolation.
We also derive $\gamma^*_m$ at the lower bound of the fixed point coupling ($g_{\mathrm{TPL}}^{*2}=2.475$) and the upper bound of the coupling ($g_{\mathrm{TPL}}^{*2}=2.823$) to include the systematic uncertainty coming from the value of the fixed point coupling.
Here, we estimate the lower bound of $g_{\mathrm{TPL}}^{*2}$ by adding the statistical and systematic errors in Eq.~(\ref{eq:TPL-g-value}) in quadrature.
We also show the results of the $s=2$ step scaling and the dependence on the scheme parameter $r$.

\subsection{Result of $s=1.5$ step scaling function in $r=1/2$ scheme}\label{sec:step-scaling-best}
First, we compute the mass anomalous dimension at the IRFP using the $s=1.5$ step scaling.
\begin{figure}[h]
\vspace{-10pt}
\begin{center}
  \includegraphics[width=12cm,angle=0]{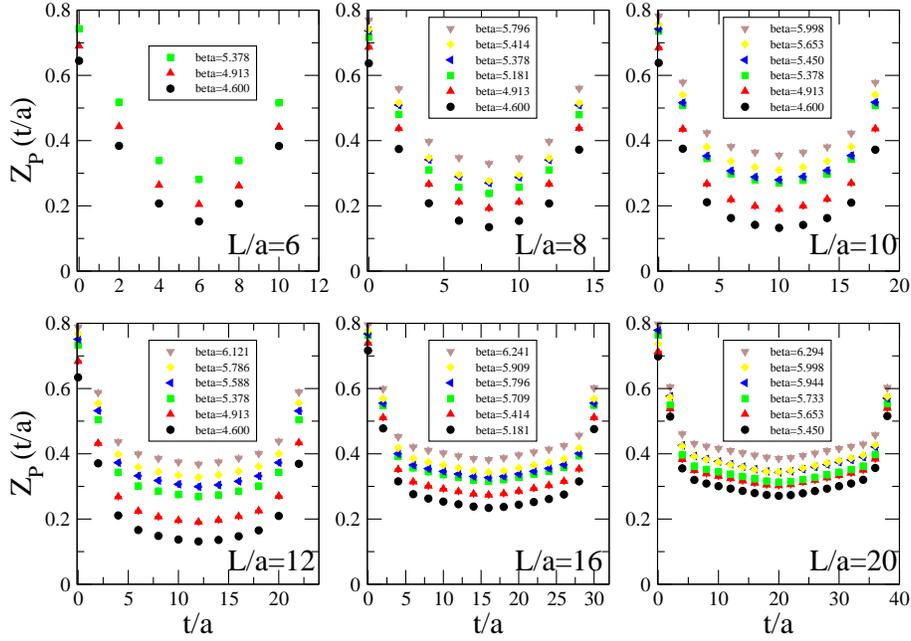}
\vspace{10pt}  
  \caption{The renormalizatioin factor $Z_P(t/a)$ for each $L/a$ and $\beta$.}
  \label{fig:raw-data-Zp}
\end{center}
\end{figure}
We show the renormalization factor $Z_P(\beta,a/L,t/a)$ in our scheme in Fig.~\ref{fig:raw-data-Zp}, and give the raw data of the factor $Z_P(t/a)$ for each lattice setup in Tables~\ref{table:nonpert-Z-L6} --~\ref{table:nonpert-Z-L20} in Appendix~\ref{sec:app-raw-data}.
We found that each $Z_P(t/a)$ has a different slope between the long $t/a$ and short $t/a$ regions.
That means that the contributing effective mass depends on the distance. 
The data in the short propagation length ($t/a \leq 6 $) for each lattice size seem to have a large discretization effect.
To reduce the effect, we choose $r=t/T=1/2$ scheme in our central analysis.

To carry out the $s=1.5$ step scaling, the interpolations of the data in $\beta$ and $L/a$ are necessary.
\begin{figure}[h]
\begin{center}
  \includegraphics[width=7cm,angle=0]{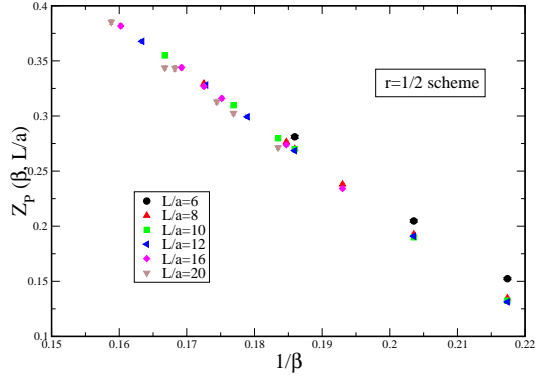}
  \caption{The $\beta$ dependence of $Z_P (\beta,L/a)$ at $t/a=T/2a$.}
  \label{fig:beta-deps-Zp-r=2}
\end{center}
\end{figure}
Figure~\ref{fig:beta-deps-Zp-r=2} shows the $\beta$ dependence of $Z_P(\beta, L/a)$ at $t/a=T/2a$.
We find that the $\beta$ dependence is approximately linear function in $1/\beta$ in the range we are considering. 
We fit four data around the value of $\beta$ needed for step scaling analysis using the fit function $Z_P(\beta)= c_0 +\frac{c_1}{\beta}$,
where $c_0$ and $c_1$ are the fitting parameters.

\begin{figure}[h]
\begin{center}
  \includegraphics[width=7cm,angle=0]{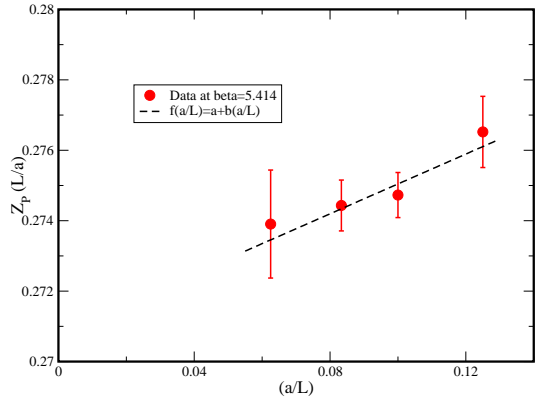}
  \vspace{1cm}
  \caption{The $a/L$ dependence of $Z_P (\beta,L/a)$ at $\beta=5.414$.}
  \label{fig:L-deps-Zp-r=2}
\end{center}
\end{figure}

The $L/a$ interpolation is simply carried out by the function $Z_P(L/a)=c_0 + c_1\frac{a}{L}$.
Figure~\ref{fig:L-deps-Zp-r=2} shows $a/L$ dependence at $\beta=5.414$ and the linear fit function in $a/L$ in the range of $1/16 \le a/L \le 1/8$.
Practically, we perform the interpolation using two lattice size data at each fixed value of $\beta$, since we have the limited number of data for each $\beta$ value.
We also carry out the other interpolation using the function $Z_P(L/a)=c_0 +c_1 \frac{L}{a}$, but the effect of the different choices of the interpolation function is negligible.

\begin{figure}[ht]
\vspace{-10pt}
\begin{center}
   \includegraphics[width=7cm]{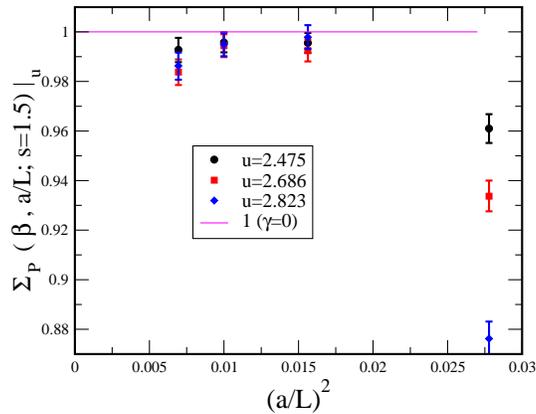}
\vspace{10pt}
 \caption{The mass step scaling function with the scheme parameter $r=1/2$ and the step scaling parameter $s=1.5$.}
 \label{fig:cont-r=2-s=1.5}
\end{center}
\end{figure}

Finally, Fig.~\ref{fig:cont-r=2-s=1.5} shows the mass step scaling function $\Sigma(\beta,a/L;s=1.5)$ on the lattice with the scheme parameter $r=1/2$.
We find that the $L/a=6$ data suffers from a larger discretization error.
The discretization effects arise from two sources.
One is a discretization effect of the renormalized coupling due to the tuned value of $\beta$.
The other comes from a discretization effect of the pseudo scalar correlator.
As we shown in Fig.~\ref{fig:raw-data-Zp}, since the data in the short propagation range ($t/a \le 6$) seems to have a large discretization, the latter is the dominant source of the large scaling violation.

Since the fit including the data at $L/a=6$ has a large chi-square, we drop the data at $L/a=6$ from the continuum extrapolation \footnote{The same situation happened in the previous work for the running coupling constant~\cite{TPL}. At that time, since $L/a=4$ data of $g_{\mathrm{TPL}}^2$ on the lattice suffer from large discretization effects, we dropped the data from the continuum extrapolation. }.
\begin{figure}[h]
\begin{center}
   \includegraphics[width=16cm]{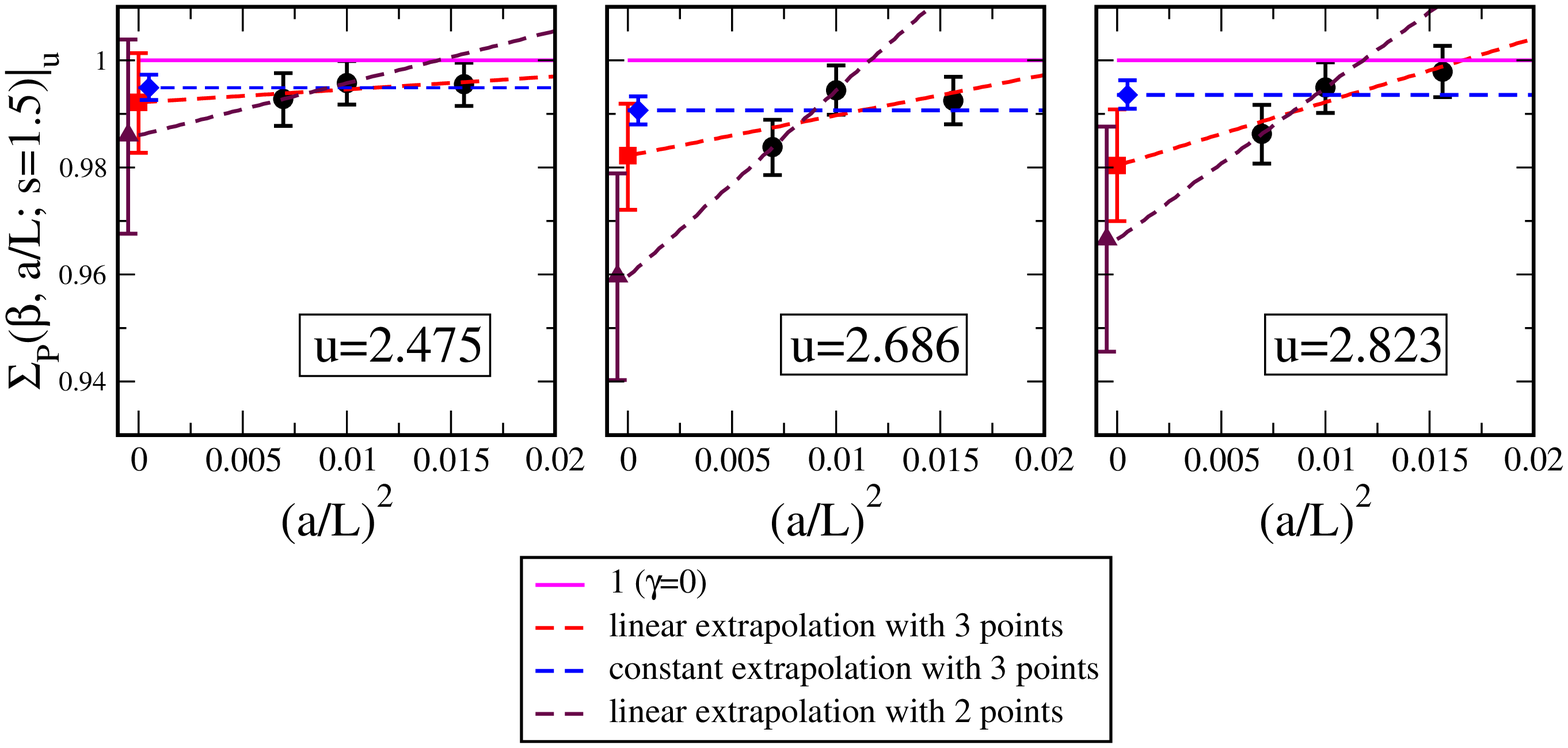}
 \caption{The continuum extrapolation of the mass step scaling function in $r=1/2$ scheme with $s=1.5$ without $L/a=6$ data. The solid line denotes the unity, where the anomalous dimension is zero. Each red and blue dashed lines denote the three-point linear and three-point constant extrapolations in $(a/L)^2$ respectively. The violet dashed line denotes the two-point linear function in $(a/L)^2$ using the finer two lattices.}
 \label{fig:cont-r=2-s=1.5-each-u}
\end{center}
\end{figure}
Figure~\ref{fig:cont-r=2-s=1.5-each-u} shows the finer three lattice data and several continuum extrapolation functions.
As a central analysis, we take the three-point linear extrapolation in $(a/L)^2$, which is drawn in red dashed line for each panels.
The $u$ dependence is small and the result is consistent with each other.
As a central result, we take the result for $u=2.686$ as a fixed point coupling.
The mass anomalous dimension of the central result with the statistical error from three-point linear fit at $u=2.686$ is given 
\beq
\gamma_m^*&=& 0.044 \hspace{3pt}_{-0.024}^{+0.025} (\mbox{stat.}) .\label{eq:result-center}
\eeq

We also carry out two different kinds of extrapolation with $\pm 1$ degree of freedom fits to estimate the systematic uncertainty of this procedure.
One is the three-point constant extrapolation (the blue dashed line in Fig.~\ref{fig:cont-r=2-s=1.5-each-u}) and the other is the two-point linear extrapolation (the violet dashed line in Fig.~\ref{fig:cont-r=2-s=1.5-each-u}).
The smallest value of $\gamma_m^\ast$ is given by the three-point constant extrapolation for $u=2.475$, and the largest one is given by the two-point linear extrapolation for $u=2.686$.
Each value of $\gamma_m^\ast$ is $0.013$ and $0.102$ respectively.
We estimate the systematic uncertainties by taking the difference between the central value and smallest or largest value respectively.
Finally, we obtain the mass anomalous dimension 
\beq
\gamma_m^*&=& 0.044 \hspace{5pt}_{-0.024}^{+0.025} (\mbox{stat.})  \hspace{5pt}_{-0.032}^{+0.057} (\mbox{syst.}) , \label{eq:final-gamma}
\eeq
where the systematic error includes the uncertainties coming from the several continuum extrapolating functions and of the choice of $u^\ast$ shown in Fig.~\ref{fig:cont-r=2-s=1.5-each-u}.

Note that the corresponding degrees of freedom for each three-point constant, three-point linear and two-point linear extrapolations are $2,1$ and $0$ respectively.
The extrapolation with the small degree of freedom might strongly suffer from the statistical fluctuation.
Furthermore, there is a signal that the finer lattice data would give a large discrepancy from the unity line, thus it might give a large anomalous dimension.
The further study including the larger lattices is necessary to give a conclusive result.

\subsection{Step scaling parameter ($s$) dependence}
We also show the result of the step scaling with $s=2$.
The advantage of the $s=2$ step scaling is that we do not need interpolations in $\beta$ and $L/a$, and the signal of the growth ratio of the factor $Z$ becomes clear.
On the other hand, the disadvantage of this is that we need the large lattice setup with the fixed distance from the continuum limit.
These differences are just technical matters, and we can make the consistency check when we carry out the step scaling with several values of $s$.
Note that the anomalous dimension within the same renormalization scheme is independent of $s$, although the mass step scaling function depends on the choice of $s$.
\begin{figure}[h]
\begin{center}
   \includegraphics[width=8cm]{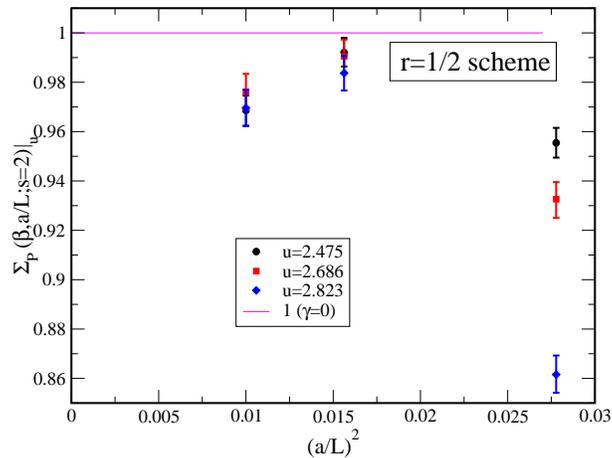}
 \caption{The mass step scaling function with the scheme parameter $r=1/2$ and the step scaling parameter $s=2$.}
 \label{fig:cont-r=2-s=2}
\end{center}
\end{figure}

Figure~\ref{fig:cont-r=2-s=2} shows the mass step scaling function in the case of $s=2$ in $r=1/2$ scheme.
Again we find that there is a large discretization error in $L/a=6$, therefore we drop the data.
The remained data points are only two, so that we take the average of these two points for each input $u$.
The degree of freedom of the continuum extrapolation is the same with the central analysis in the $s=1.5$ step scaling analysis. 
The anomalous dimension is obtained 
$\gamma_m^*= 0.028 \pm{0.006}$ for $u=2.475$,
$\gamma_m^*= 0.023\pm {0.007}$ for $u=2.686$ and
$\gamma_m^*= 0.034 \hspace{2pt}_{- 0.008}^{+0.007}$ for $u=2.823$.
These results are consistent with the ones of $s=1.5$ analysis within $1\sigma_{\rm stat.}$.
That is an indirect check for the effects of the $\beta$ and $L/a$ interpolations in the $s=1.5$ step scaling.

\subsection{Scheme parameter ($r$) dependence}\label{sec:r-dependence}
We also show the result in the scheme with $r=1/3$.
Changing $r$ corresponds to the change of renormalization condition, so that it gives a different renormalization scheme with the same lattice setup.
If the theory is not at the fixed point, the value of the anomalous dimension depends on the renormalization scheme, namely the choice of $r$.
However at the fixed point, it should be independent of the choice of the renormalization scheme. 

\begin{figure}[h]
\vspace{1cm}
\begin{center}
   \includegraphics[width=14cm]{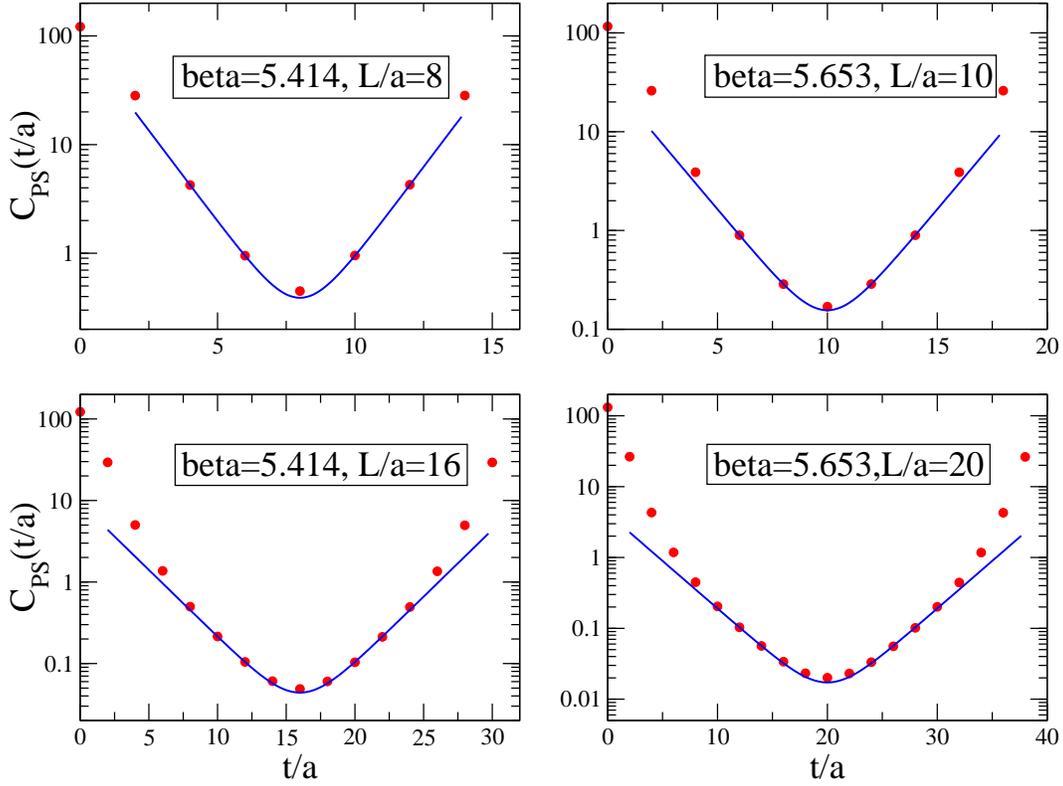}
 \caption{The interpolation of the nonperturbative correlation function of the pseudo scalar operator in $t/a$ for each lattice size for $r=1/3$ scheme. For $L/a=6$ and $L/a=12$, we do not need the interpolation.}
 \label{fig:Cp-int-t}
\end{center}
\end{figure}

In $r=1/3$ scheme how to estimate the value of correlator at non-integer $t/a$ in several lattice sizes is a problem.
As discussed in Appendix~\ref{sec:shape-correlation}, we expect that the correlation function in the finite box can be described by the exponential functions even though the theory is conformal.
Since we impose the periodic boundary condition in the temporal direction, the correlator is proportional to the linear combination of the cosh function of several energy modes, $\sum_{i} C_i (\omega_i)  \cosh[\omega_i (\hat{t}-\hat{T}/2)]$. 
Here $\omega_i$ is the effective mass for each mode.
We assume that our data can be fitted by a single $\cosh$ function in the small range of $\hat{t}$.
Figure~\ref{fig:Cp-int-t} shows the examples of the $\hat{t}$ interpolation for $\hat{t}=\hat{T}/3$.
We take the following fit range for each lattice size:
$(4 \le t/a \le 6)$ and $(10 \le t/a \le 12)$ for $L/a=8$, $(6 \le t/a \le 8)$ and $(12 \le t/a \le 14)$ for $L/a=10$, $(10 \le t/a \le 12)$ and $(20 \le t/a \le 22)$ for $L/a=16$ and $(12 \le t/a \le 14)$ and $(26 \le t/a \le 28)$ for $L/a=20$.
The blue curve in Fig.~\ref{fig:Cp-int-t} denotes the fit function.
Note that essentially the number of independent data points is two, since there is periodicity in temporal direction and the staggered fermion takes value only on even-site. 
Thus we solve the equation $C_{PS}(\hat{t})=a\cosh(b(\hat{t}-\hat{T}/2))$ to determine the fit parameters ($a$ and $b$).

\begin{figure}[h]
\begin{center}
   \includegraphics[width=8cm]{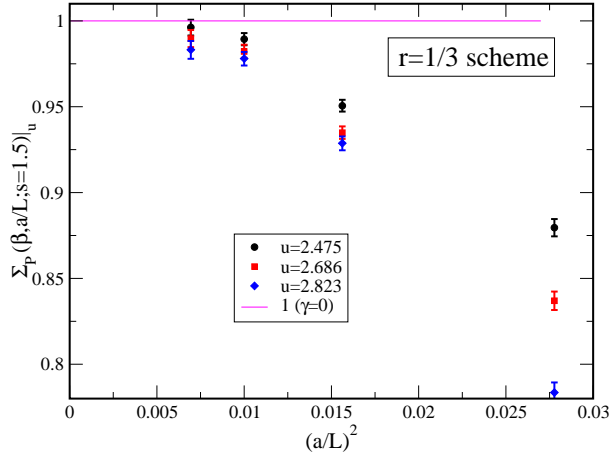}
 \caption{The mass step scaling function with the scheme parameter $r=1/3$ and the step scaling parameter $s=1.5$.}
 \label{fig:cont-r=3-s=1.5}
\end{center}
\end{figure}

We also carry out the same interpolation for the tree level correlators, and repeat the same analysis as we shown in Sec.~\ref{sec:step-scaling-best}.
Figure~\ref{fig:cont-r=3-s=1.5} shows the mass step scaling function for each input value of $u$.
We find that there are the large scale violations in particular $L/a=6$ and $8$ data. 
As we explained, the origin of these scale violations comes from two sources: the discretization error of the input renormalized coupling constant and the one of the correlation function.
Both Fig.~\ref{fig:cont-r=2-s=1.5} and Fig.~\ref{fig:cont-r=3-s=1.5} have the same discretization error coming from the scale violation of the coupling constant.
The difference of $\Sigma(\beta,a/L)$ between these two plots for each data point shows the discretization error of the pseudo scalar correlator.
The data at $L/a=6$ and $L/a=8$ include the data of $Z_P(t/a)$ at $t/a \le 6$ in Fig.~\ref{fig:raw-data-Zp}.
We expect that such data in the short propagation length gives the large scaling violation.

We estimate the value of step scaling function in the continuum limit using the constant extrapolation for two finer lattice data.
Thus the degree of freedom of the continuum extrapolation is the same with the central analysis in $r=1/2$ scheme again. 
The anomalous dimension is given by
$\gamma_m^*= 0.020 \pm{0.007} $ for $u=2.475$,
$\gamma_m^*= 0.037\pm {0.008}$ for $u=2.686$ and
$\gamma_m^*= 0.050 \pm{0.008}$ for $u=2.823$.
These results are also consistent with the result (\ref{eq:final-gamma}) within $1\sigma_{\rm stat.}$.
This is an evidence that the anomalous dimension at the IRFP shows the universal property.

\section{Discussion}\label{sec:discussion}
Our result of the mass anomalous dimension at the IRFP is  
\beq
 \gamma_m^*&=& 0.044 \hspace{5pt}_{-0.024}^{+0.025} (\mbox{stat.}) _{-0.032}^{+0.057} (\mbox{syst.}), 
\eeq
where the systematic error includes the uncertainty of both continuum extrapolations and $u^\ast$ dependence.

\begin{figure}[h]
\begin{center}
   \includegraphics[width=10cm]{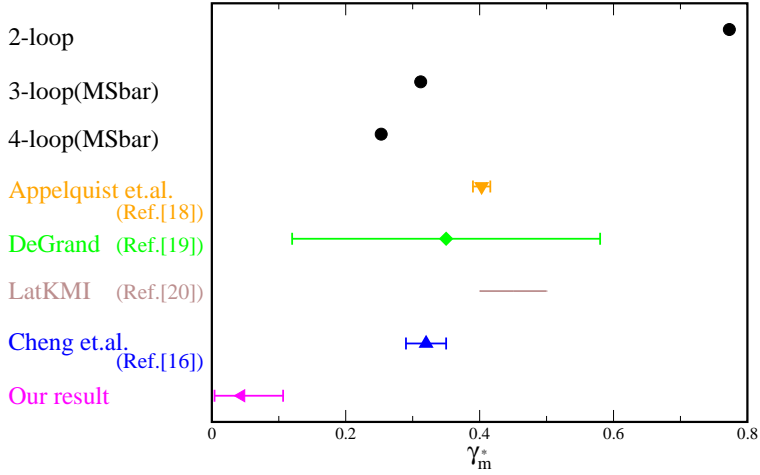}
   \vspace{1.5cm}
 \caption{The comparison of the mass anomalous dimension at IRFP for several studies. From the top, the perturbative 2-loop result, 3-loop $\overline{{\mathrm{MS}}}$, 4-loop $\overline{{\mathrm{MS}}}$, the recent lattice results in the papers~\cite{Appelquist:2011dp}, \cite{DeGrand:2011cu},  \cite{Aoki:2012eq}, \cite{Cheng:2013eu} and our result. Note that in the papers~\cite{DeGrand:2011cu, Aoki:2012eq}  there is no `` $^*$ " on the gamma in their own papers.}
  \label{fig:comp-results}
\end{center}
\end{figure}

Let us compare our result with the other predictions.
Figure~\ref{fig:comp-results} shows the values of the mass anomalous dimension in other literatures.
The perturbative results \cite{Vermaseren:1997fq, Ryttov:2010iz,Mojaza:2010cm} give $\gamma_m^*=0.773$ in the 2-loop and $\gamma_m^{* }=0.312, 0.253$ in $3$- and $4$-loop in $\overline{{\mathrm{MS}}}$ scheme.
The value of fixed point coupling is $\alpha^*=0.75$ in $2$-loop analysis, so that the perturbative expansion does not show a good convergence.
On the other hand, in the case of $N_f=16$, the IRFP exists in the perturbative region, and the anomalous dimension is predicted $\gamma_m^*=0.0272, 0.0258$ and $0.0259$ in $2$-loop, $3$-loop and $4$-loop respectively.
The perturbative results for $N_f=16$ shows a good convergence around $\gamma_m^\ast \sim 0.026$ and the value seems to be reliable.
We expect that the value of $\gamma_m^\ast$ monotonically increases when $N_f$ is decreasing in the conformal window, so that the value of $\gamma_m^\ast$ for $N_f=16$ case gives the lower bound.
Our result is quite smaller than the perturbative prediction in the case of $N_f=12$, but it is the same order with $N_f=16$ case.

Furthermore, it is also important to compare our result with the other nonperturbative lattice studies.
Before presenting detailed discussion, we would like to list criteria for reliable studies of the IRFP in QCD-like theories using the lattice regularization.
\begin{itemize}
\item Keep in the asymptotic free region.
\item Take the continuum limit ($a \rightarrow 0$) keeping a physical input constant.
\item Take the infinite volume limit ($L \rightarrow \infty$).
\item Take the chiral limit ($m \rightarrow 0$).
\end{itemize}
In our works, firstly, we already show that $g_{\mathrm{TPL}}^{*2}=2.686$ is the first zero of the beta function from the perturtabative region (Fig.~15 in Ref.~\cite{TPL}), so that the lattice data, where we  are investigating, connect to the asymptotic free region.
Secondly, we have taken the continuum limit. The important point is that we have to keep an input physical quantity constant to carrying out this procedure, since if we take naively the $\beta \rightarrow \infty$ limit then the theory becomes the UV gaussian fixed point.
In our works, we fix the renormalized coupling, which is essentially the correlator of the Polyakov loop in the TPL scheme, as an input to taking the continuum limit.
Thirdly, since the running coupling constant at $\mu =1/L$ stop growing when the scale is changed $L \rightarrow 2L$ shown in Ref.~\cite{TPL}, we can regard that the scale is enough infrared to reach at the vicinity of the infrared fixed point.
Fourthly, we can exactly carry out the massless simulation thank to the twisted boundary condition.

On the other hand, there are four papers to estimate the mass anomalous dimension in this theory using the mass deformation method.
The one of closest results, $\gamma_m^\ast=0.32(3)$, is given by Cheng {\it et. al.} in the paper~\cite{Cheng:2013eu}.
The hyperscaling (volume-scaling) of the Dirac eigenmodes for several values of $\beta$ has been investigated in the (approximately) massless limit.
The results show a reliable scaling behavior and they can estimate both the fixed point value of $\beta$ in their lattice gauge action and the universal mass anomalous dimension in IR limit.
Although they have studied wide range of $\beta$ and roughly determine the critical fixed point of the $\beta$, the second procedure of the list, namely taking the continuum limit {\it{with keeping a physical quantity at the scale}}, has not been carried out.
The discrepancy between our result and their result might come from an insufficient estimation of the systematic uncertainty coming from the continuum extrapolation in our analysis or the insufficiency of the tuning of $\beta$ in hyperscaling study.

In the papers~\cite{Appelquist:2011dp} and  \cite{DeGrand:2011cu} the authors use a part of the data in the paper \cite{Fodor:2011tu}, and fit these data using the hyperscaling ansatz~\cite{Miransky:1998dh} on the lattice.
Orginally, the paper \cite{Fodor:2011tu} shows that the data can be fitted by the weakly chiral symmetry broken hypothesis better than the conformal hypothesis. 
However, the paper \cite{Appelquist:2011dp} shows if we fit only the largest lattice size data to avoid serious finite volume effects, the conformal hypothesis also works well.
The universal fit using the hyperscaling for several hadronic spectrum gives the mass anomalous dimension: $\gamma_m^*=0.403(13)$.
The paper \cite{DeGrand:2011cu} shows the finite-size scaling using the same data, and the anomalous dimension of the pseudo scalar operator from the mass spectrum is given by $\gamma_m=0.35(23)$.
In the paper \cite{Aoki:2012eq} by LatKMI  collaboration, they also use the same method, and fit their own data. 
The anomalous dimension in paper~\cite{Aoki:2012eq} is given as $\gamma_m \sim 0.4$--$0.5$.
Here, we should mention that in the papers~\cite{DeGrand:2011cu,Aoki:2012eq} the symbol `` $^\ast$ " is not added to the anomalous dimension, which denotes a symbol of the quantities at the IRFP.
Actually, LatKMI also has been studying the mass anomalous dimension in the case of $N_f=16$ fermions using the same strategy~\cite{Aoki:2012yd}.
However, the result of the paper~\cite{Aoki:2012yd} using the hyperscaling also shows a scaling behavior within the wide range of the larger mass anomalous dimension, and do not show the perturbative converged value of the mass anomalous dimension $\gamma_m^\ast \sim 0.026$.
If it would not be the anomalous dimension at the fixed point, then it depends on the renormalization scheme and the discrepancy is not a problem.

In fact, the SU($2$) gauge theory coupled to $N_f=2$ adjoint fermion is also known as an IR conformal field theory.
Several independent collaborations have been deriving the mass anomalous dimension at the IRFP.
The step scaling method using the SF scheme gives the predictions $0.05 \le \gamma_m^\ast \le 0.56$ in paper~\cite{Bursa:2009we} and $\gamma_m^\ast=0.31(6)$ in paper~\cite{DeGrand:2011qd} respectively.
The hyperscaling for the string tension, Meson spectrum and the mode number of the Dirac operator give $\gamma_m^\ast=0.22(6)$ (Ref.~\cite{DelDebbio:2010hx}), $0.05 \le \gamma_m^\ast \le 0.20$ (Ref.~\cite{DelDebbio:2010hu}) and $\gamma_m^\ast=0.51(16)$ (Ref.~\cite{Giedt:2012rj}), and $\gamma_m^\ast=0.371(20)$ (Ref.~\cite{Patella:2012da}).
There are somewhat consistent with each other, while some values have a large errorbar and there are also unknown systematic errors.
The paper \cite{Bursa:2009we} determined the critical value of $\beta$ around $\beta \sim 2.25$ using the step scaling method in the SF scheme. 
Also in the paper~\cite{Giedt:2011kz}, the Creutz ratio does not run around $\beta=2.25$ and that is a signal of the fixed point of the Wilson loop coupling
\footnote{Generally the value of $\beta$ at the fixed point can depend on the lattice gauge action or the boundary conditions, since the different lattice setup gives a different discretization errors. There was no guarantee that the lattice gauge action with SF boundary condition in paper~\cite{Bursa:2009we} gives the same critical value of $\beta$ at the IRFP with the periodic boundary condition in the paper~\cite{Giedt:2011kz}. }.
Then in the papers~\cite{DelDebbio:2010hx} -- \cite{Giedt:2012rj}, they derive the mass anomalous dimension using the hyperscaling with tuned value of $\beta=2.25$.
We consider that such tuned value of $\beta$ to realize the IRFP is necessary to obtain the universal anomalous dimension using the hyperscaling for the mass deformed gauge theory.

\section{Summary }\label{sec:summary}
We propose a new renormalization scheme for the composite operators.
In this renormalization scheme, the correlator of the pseudo scalar operator satisfies the ``tree level renormalization condition" in coordinate space, in which the renormalized value is equal to the tree level amplitude at the fixed propagation length.
We introduce the twisted boundary conditions for the spatial directions. That makes us to obtain the tree level correlator for the massless fermions.
In this scheme, the different propagation length corresponds to the different renormalization schemes, and we choose the suitable length in practical simulations.

Furthermore we study the mass step scaling function for the SU($3$) $N_f=12$ massless gauge theory using this renormalization scheme.
Using the PCAC relation, the mass renormalization factor is related to the renormalization factor of the pseudo scalar operator.
We actually measure the renormalization factor of the pseudo scalar operator, and directly derive the mass anomalous dimension.
This work is the first study of the derivation of the mass anomalous dimension at the IRFP using the step scaling method.

Our result of the mass anomalous dimension at the IRFP of this theory with the scheme parameter $r=1/2$ and the step scaling size $s=1.5$ is   
\beq
 \gamma_m^*&=& 0.044 \hspace{5pt}_{-0.024}^{+0.025} (\mbox{stat.}) _{-0.032}^{+0.057} (\mbox{syst.}) ,
\eeq
where the systematic error includes the uncertainty of both continuum extrapolations and $u^\ast$ dependence.
We also investigate the step scaling parameter ($s$) dependence and the scheme parameter ($r$) dependence of $\gamma^*_m$.
The results with the different choices of $s$ and $r$ are consistent with each other within $1\sigma_{\rm stat.}$ discrepancy.
Note that in the current analysis the continuum extrapolation is done with small degrees of freedom, and the result might be strongly affected by the statistical fluctuation.
Furthermore, there is a signal that the finer lattice data would give a large discrepancy from the unity line, thus it might give a large anomalous dimension.
Further careful estimation of the systematic uncertainty from the continuum extrapolation might be important.
We will report the conclusive results including the larger lattice simulation in the forthcoming paper.

\begin{figure}[h]
\vspace{1cm}
\begin{center}
  \includegraphics[width=6cm]{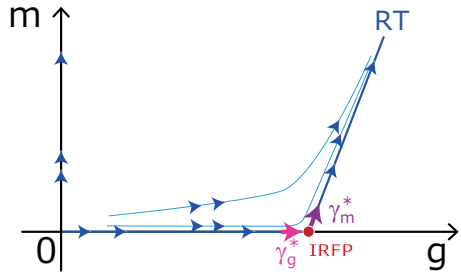}
 \caption{The theory space for the SU($3$) $N_f=12$ gauge theory.}
 \label{fig:theory-space}
\end{center}
\end{figure}

If there is no other relevant operator, then the renormalization group flows of the SU($3$) $N_f=12$ gauge theory are governed by the two dimensional theory spaces whose coordinates are the fermion mass and the gauge coupling constant (See: Fig.~\ref{fig:theory-space}).
The universal quantities to characterize the IRFP are the critical exponent of the beta function ($\gamma_g^*$) and the mass anomalous dimension ($\gamma_m^*$).
We have investigated the renormalization group flow on the massless line. 
We also derived
\beq
\gamma_g^\ast = 0.57^{+0.35}_{-0.31} (\mbox{stat.}) _{-0.16}^{+0}\, (\mbox{syst.}).
\eeq
in our previous work~\cite{TPL}.
We determine $\gamma_g^\ast$ and $\gamma_m^\ast$ using the TPL scheme for renormalized coupling constant and the new scheme for the fermion mass respectively.
Changing the renormalization schemes corresponds to the coordinate transformation of the theory spaces.
The existence of IRFP is independent object to the coordinate transformation. 
The values of $\gamma_g^\ast$ and $\gamma_m^\ast$ are also universal, since they are the eigenvalues of two linearized $\beta$ functions around the IRFP.

We compare our result with other lattice studies.
All other studies has been done based on the scaling law.
There is a large difference between our result and their, but some works utilize the mass deformed theory without the tuning of $\beta$.
We consider that an insufficient parameter tuning for the hyperscaling could be a reason of the discrepancy.
The determination of $\gamma_m^\ast$ using the hyperscaling method works well only if the renormalization group flow reach as the vicinity of the IRFP such as the solid curve in Fig.~\ref{fig:theory-space}.
Since we do not know the action at the IRFP, we introduce the fermion mass term in the lattice gauge action at the UV gaussian fixed point.
Around the gaussian fixed point, the mass term is relevant operator and the gauge coupling is marginal, so that generally the renormalization group flow goes away from the massless axis such as the dotted curve in Fig.~\ref{fig:theory-space}.
If it happens, the renormalization group flow reaches the renormalized trajectory (RT in Fig.~\ref{fig:theory-space}) where it is far away from the IRFP.
The anomalous dimension changes along the renormalization group flow even on the RT.
If there is no scale invariance and the mass anomalous dimension is not the one at the IRFP, then generally the value of $\gamma_m$ depends on the renormalization scheme.

To find both the IRFP and to obtain the universal mass anomalous dimension needs two independent observables.
It is impossible to do both only using the hyperscaling for the mass deformed gauge theory in two parameter spaces ($\beta, m$).
The hyperscaling for the mass deformed conformal gauge theory is a powerful method to obtain the precise value of the anomalous dimension.
However, we would like to emphasize the tuning of the lattice parameters ($\beta, L/a$ and the fermion bare mass) in order to realize the vicinity of the IRFP is important to obtain the universal quantity.

The future direction within our work is to carry out the simulation with the larger lattice size ($(L/a)^3 \times (T/a)=24^3 \times 48$) to give a conclusive value of the critical exponent at the IRFP.
Actually, in the present analysis the degree of freedom of the continuum extrapolation is only one, so that we did not estimate the systematic uncertainty coming from this procedure.
We will report our final results including further large lattice data in forthcoming paper.
To measure the wave function renormalization factor for other hadronic operators and to investigate the universal scaling behavior are also interesting.
Furthermore, if we take the continuum limit carefully, then a study with the different lattice setup using the tuned values of $\beta$ to realize $g_{\mathrm{TPL}}^{\ast 2}$ (Table~\ref{table:beta-L}) would be promising to derive the anomalous dimension at the IRFP.
That must be a nontrivial check for the universality using the lattice simulations
\footnote{We should take care to avoiding the artifact phase or the strong coupling phase reported in~\cite{Cheng:2011ic, Deuzeman:2012ee, deForcrand:2012vh}. }.

\section*{Acknowledgements}
The idea of this novel scheme was suggested to us by T.~Onogi.
We would like to thank him and H.~Ikeda for useful discussions and Y.~Taniguchi for providing us with notes on the free fermion correlator in the infinite volume in Appendix~\ref{sec:shape-correlation}.
The simulation codes was originally developed by H.~Matsufuru in our previous works, we also thank him.
We thank S.~Hashimoto, Y.~Iwasaki, M.~L\"{u}scher, A.~Patella, F.~Sannino, H.~Suzuki, N.~Yamada and S.~Yamaguchi for giving useful comments and discussions.
We also would like to thank A.~Irie for making Figs.~\ref{fig:strategy} and \ref{fig:theory-space}.
Numerical simulation was carried out on
Hitachi SR16000 at YITP, Kyoto University,
NEC SX-8R at RCNP, Osaka University,
and Hitachi SR16000 and IBM System Blue Gene Solution at KEK 
under its Large-Scale Simulation Program
(No.~12/13-16), as well as on the GPU cluster at Osaka University.
We acknowledge Japan Lattice Data Grid for data
transfer and storage.
E.I. is supported in part by
Strategic Programs for Innovative Research (SPIRE) Field 5.


\appendix

\section{The shape of the correlation function}\label{sec:shape-correlation}
We give a comment on the functional form of the pseudo scalar correlator.
If the theory is conformal, then in the infinite volume the correlation function of an operator $\mathcal{O}$ shows the power function,
\beq
 \langle {\mathcal O} (t,\vec{x}) {\mathcal O} (0,\vec{0}) \rangle=\frac{const.}{|t|^{2\Delta_{\mathcal O}}},
\eeq
where $\Delta_{\mathcal O}$ denotes the conformal dimension of the operator, which is given by the sum of the canonical and anomalous dimensions of the operator respectively.
However, it is hard to fit the lattice data using the power function.
We consider that there are two reasons why the data in our simulation shows cosh behavior not power law.

First, we consider that the free massive fermion theory.
In the continuum limit with infinite volume, we can calculate the correlation function of the pseudo scalar operator in this theory.
Let us consider the correlation function of the pseudo scalar operator ($P$).
The correlation function can be obtained using two free fermion propagators
\beq
G(t,m)&=& \int d^3 \vec{x} \langle P (t,\vec{x}) P (0,\vec{0}) \rangle, \nonumber\\
&=&\int d^3 \vec{x} \int \frac{d^4 p}{(2\pi)^4} \int \frac{d^4 k}{(2\pi)^4} \frac{\mbox{tr} (p_\mu \gamma^\mu k_\nu \gamma^\nu +m^2)}{(p^2+m^2)(k^2+m^2)} e^{i(p-k)x}\nonumber\\
&=& \frac{1}{\pi^2} \int^\infty_0 d \tilde{p} \tilde{p}^2 \exp(-2 \sqrt{\tilde{p}^2+m^2} t),
\eeq
where $p,k$ denote four-dimensional momenta and $\tilde{p}$ denotes a magnitude of three-dimensional momentum.
In the case of the massive fermion, it is written by
\beq
G(t,m)=\frac{1}{\pi^2}\frac{m^2}{2t} K_2(2mt),
\eeq
where $K_2(2mt)$ is the modified Bessel function.
In the case of the massless fermion, the theory is a conformal free theory.
The correlation function is given by
\beq
G(t,m=0)=\frac{1}{4\pi^2 t^3}, \label{eq:t-cubic}
\eeq
as expected by the dimensional analysis.
\begin{figure}[h]
\begin{center}
   \includegraphics[width=15cm]{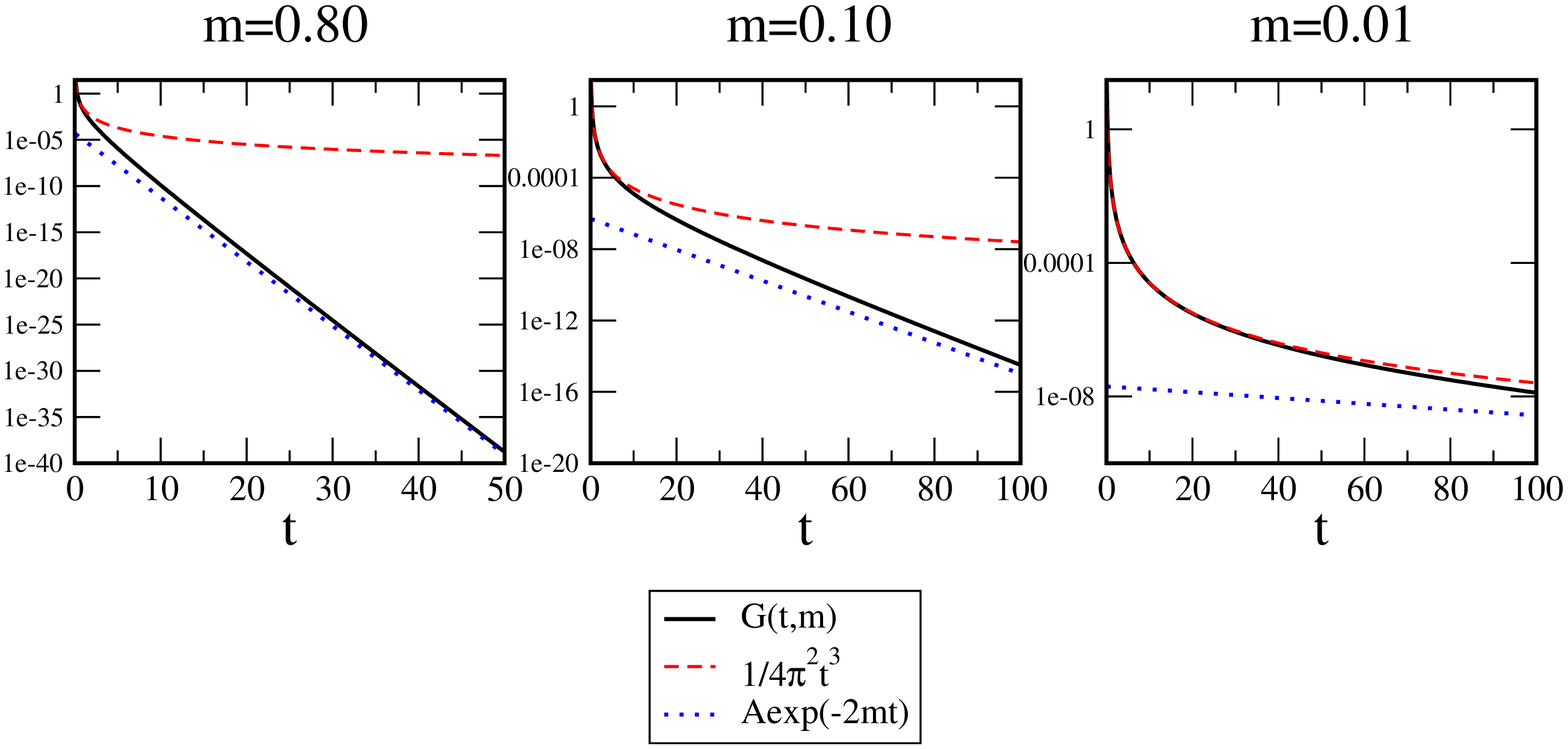}
 \caption{The mass dependence of the correlation function at tree level in the infinite volume. The vertical axis is log-scale. The dashed (red) curve and dot (blue) line denote the massless correlation function Eq.~(\ref{eq:t-cubic}) and the exponential decay function respectively. The coefficient $A$ of the dot line is chosen by hand in the plots. The comparison only the slopes between $G(t,m)$ and $A \exp(-2mt)$ makes sense. }
 \label{fig:bessel}
\end{center}
\end{figure}

Figure~\ref{fig:bessel} shows the shape of the $G(t,m)$ with the mass $m=0.80, 0.10, 0.01$.
If the mass is large, then the correlation function $G(t,m)$ can be described by $e^{-2mt}$.
On the other hand, if the mass is a quite small, the $G(t,m)$ reproduces the massless power correlation function.
In the middle range of the mass ($m \sim 0.1$), the correlation function can be described by the power function (Eq.~(\ref{eq:t-cubic})) in the short $t$ range,
while it is consistent with the exponential function $e^{-2mt}$ in the long $t$ region.
If the mass becomes small, the available range of the power function fit becomes broad.
That is related to the fact that the massless free fermion theory is an UV fixed point, so that a short range behavior describes a conformal behavior. 

On the lattice, since we introduce other two scales: the finite lattice size and a lattice spacing, the discussion becomes complicated.
In our simulation, the twisted boundary condition shifts the zero momentum to nonzero values.
The shifted momentum plays a similar role to the mass.

\begin{figure}[h]
\vspace{1cm}
\begin{center}
   \includegraphics[width=8cm]{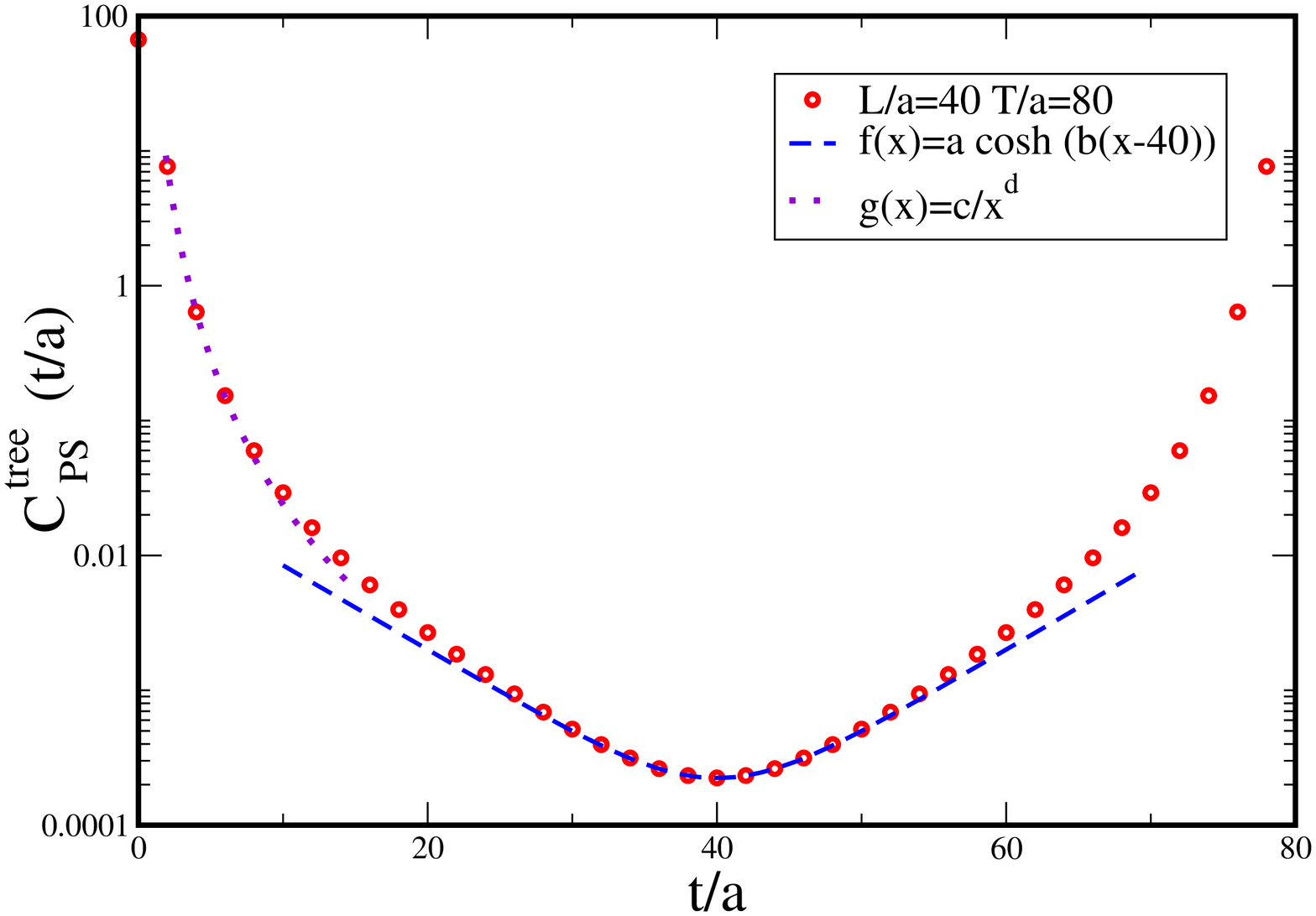}
 \caption{The correlation function at tree level for $L/a=40, T/a=80$. The function $f(x)$ denote a cosh function which is obtained by solving the equation using the data in $38 \le t/a \le 42$. The function $g(x)$ denotes a power function which is obtained by solving the equation using the data in $2 \le t/a \le 4$.}
 \label{fig:Ct-tree-L40-T80}
\end{center}
\end{figure}

Figure~\ref{fig:Ct-tree-L40-T80} shows the data of the tree level correlator in the larger lattice size ($L/a=40, T/a=80$).
In the long propagation length, the data can be fitted by a cosh function, $f(x)=a \cosh(b(x-40))$.
At that time, the effective mass, which is shown in Fig.~\ref{fig:omega0}, is $\omega \sim  0.144$.
In the short propagation length, the correlation function can be fitted by the power function, $g(t)=c/t^{d}$.
If we determine the fit parameter using the data at $t/a=2$ and $4$ by solving the equation, the exponent ($d$) becomes $3.59$.
There is a small discrepancy from $d=3$ in Eq.~(\ref{eq:t-cubic}).
We expect that it is an effect of the UV cutoff (=lattice spacing).

Figure~\ref{fig:Ct-tree-L40-T80} is qualitatively consistent with the middle panel of Fig.~\ref{fig:bessel}.
In our main analysis in this paper, we use the smaller lattice size than $L/a=40, T/a=80$, therefore the value of the effective mass ($\omega$) is larger than the value of $\omega$ in $L/a=40$.
That is the reason why the shape of the correlation functions on our lattice shows cosh behavior, while we should stress that we take the continuum limit and the fermion mass is zero at the limit.

The other reason might come from the finiteness of the lattice extent.
In two-dimensional (interactive) conformal field theory, the conformal map from the infinite plane to the cylinder with the radius ($L$) is known.
On the cylinder coordinate, the correlation function becomes exponential function if the distant (in the direction with the infinite length of the cylinder) becomes larger than the compact radius $L$ (See \cite{cft}),
\beq
\langle \phi (t,x) \phi (0,0) \rangle \sim \left( \frac{2 \pi}{L} \right)^{2\Delta_\phi}  \exp ( -\frac{2 \pi t \Delta_\phi}{L}).
\eeq
In our simulation, the temporal lattice extent is twice larger than the spatial one, and we found that the data in $t/a \ge L/2a$ regime shows the exponential behavior.

Based on these considerations, we fit our lattice data using cosh function in Sec.~\ref{sec:tree-level} and Sec.~\ref{sec:r-dependence}.

\newpage
\section{Raw data}\label{sec:app-raw-data}
\begin{table}[h]
\begin{center}
\caption{The tree level correlator within the nontrivial vacua for each lattice size. } \label{table:tree-Ct}
\begin{tabular}{|c|c|c|c|c|c|}
\hline
$L/a$  &   $ t/a=2 $ & $t/a=4$ & $t/a=6$ & $t/a=8$ & $t/a=10$ \\
\hline\hline
6          &   $7.264$ &  $4.281 \times 10^{-1}$   & $9.744 \times 10^{-2}$  & -& - \\     
\hline
8          &   $7.502$ &  $5.127 \times 10^{-1}$   & $8.261 \times 10^{-2}$  & $3.431\times 10^{-2}$  & -  \\     
\hline
10          &   $7.593$ &  $5.634 \times 10^{-1}$   & $1.016 \times 10^{-1}$  & $2.912 \times 10^{-2}$  & $1.625 \times 10^{-2}$  \\     
\hline
12          &   $7.632$ &  $5.926 \times 10^{-1}$   & $1.168 \times 10^{-1}$  & $3.496 \times 10^{-2}$  & $1.370 \times 10^{-2}$  \\     
\hline
16          &   $7.659$ &  $6.200 \times 10^{-1}$   & $1.350 \times 10^{-1}$  & $4.526 \times 10^{-2}$  & $1.834 \times 10^{-2}$  \\     
\hline
20          &   $7.668$ &  $6.306 \times 10^{-1}$   & $1.438 \times 10^{-1}$  & $5.150 \times 10^{-2}$  & $ 2.238 \times 10^{-2}$  \\     
\hline \hline
$L/a$ & $t/a=12$ & $t/a=14$ & $t/a=16$ & $t/a=18$  & $t/a=20$\\
\hline\hline
12        & $9.022 \times10^{-3}$ & - & - & - & - \\
\hline
16        & $8.463 \times10^{-3}$ & $4.664 \times 10^{-3}$ & $3.655 \times 10^{-3}$ & - & - \\
\hline
20        & $1.083 \times10^{-2}$ & $5.679 \times 10^{-3}$ & $3.252 \times 10^{-3}$  & $2.154 \times 10^{-3}$ & $1.837 \times 10^{-3}$  \\
\hline
\end{tabular}
\end{center}
\end{table}

\begin{table}[h]
\begin{center}
\caption{The Data of $Z_P$ factor ($L/a=6,T/a=12$)} \label{table:nonpert-Z-L6}
\begin{tabular}{|c|c|c|c|}
\hline
{} &   $ \beta=4.600 $ & $\beta=4.913$ & $\beta=5.378$  \\
\hline\hline
$t/a=2$    &  $0.3840 _{-19}^{+19}$ & $0.44316 _{-19}^{+20} $ & $0.5178 _{-18}^{+17}$ \\     
$t/a=4$    &  $0.2074 _{-12}^{+12}$ & $0.26362 _{-14}^{+14} $ & $0.3388 _{-16}^{+15}$ \\     
$t/a=6$    &  $0.1524 _{-09}^{+09}$ & $0.20474 _{-11}^{+11} $ & $0.2811 _{-13}^{+13}$ \\     
$t/a=8$    &  $0.2072 _{-12}^{+12}$ & $0.26157 _{-13}^{+14} $ & $0.3385 _{-15}^{+15}$ \\     
$t/a=10$   &  $0.3834 _{-18}^{+18}$ & $0.44132 _{-18}^{+18} $ & $0.5159 _{-18}^{+18}$ \\     
\hline                                  
\end{tabular}
\end{center}
\end{table}

\begin{table}[h]
\begin{center}
\caption{The Data of $Z_P$ factor ($L/a=8,T/a=16$)} 
\begin{tabular}{|c|c|c|c|c|c|c|}
\hline
{} &   $ \beta=4.600 $ & $\beta=4.913$ & $\beta=5.181$ & $\beta=5.378$ & $\beta=5.414$ & $\beta=5.796$  \\
\hline\hline
$t/a=2$  & $0.3743_{-19}^{+19}$ &  $0.4371 _{-18} ^{+18}$ &  $0.4806_{-14}^{+13}$ & $0.5088_{-18}^{+17}$ & $0.5156_{-13}^{+13}$ & $0.5585_{-12}^{+13}$  \\     
$t/a=4$  & $0.2079_{-13}^{+13}$ &  $0.2668 _{-14} ^{+14}$ &  $0.3103_{-11}^{+11}$ & $0.3414_{-16}^{+15}$ & $0.3475_{-11}^{+11}$ & $0.3968_{-11}^{+12}$  \\     
$t/a=6$  & $0.1546_{-11}^{+10}$ &  $0.2121 _{-13} ^{+12}$ &  $0.2570_{-10}^{+10}$ & $0.2894_{-15}^{+15}$ & $0.2949_{-11}^{+11}$ & $0.3472_{-11}^{+12}$  \\     
$t/a=8$  & $0.1348_{-09}^{+09}$ &  $0.1927 _{-11} ^{+11}$ &  $0.2380_{-09}^{+10}$ & $0.2702_{-15}^{+15}$ & $0.2765_{-10}^{+10}$ & $0.3295_{-11}^{+11}$  \\     
$t/a=10$ & $0.1541_{-10}^{+11}$ &  $0.2122 _{-12} ^{+12}$ &  $0.2569_{-11}^{+11}$ & $0.2876_{-17}^{+16}$ & $0.2943_{-11}^{+11}$ & $0.3465_{-11}^{+11}$  \\     
$t/a=12$ & $0.2072_{-12}^{+12}$ &  $0.2672 _{-14} ^{+14}$ &  $0.3100_{-11}^{+12}$ & $0.3408_{-16}^{+16}$ & $0.3462_{-11}^{+11}$ & $0.3970_{-11}^{+11}$  \\     
$t/a=14$ & $0.3723_{-19}^{+19}$ &  $0.4384 _{-18} ^{+19}$ &  $0.4795_{-14}^{+14}$ & $0.5088_{-17}^{+17}$ & $0.5148_{-12}^{+13}$ & $0.5599_{-11}^{+12}$  \\     
\hline                                  
\end{tabular}
\end{center}
\end{table}

\begin{table}[h]
\begin{center}
\caption{The Data of $Z_P$ factor ($L/a=10,T/a=20$)} 
\begin{tabular}{|c|c|c|c|c|c|c|}
\hline
{} &   $ \beta=4.600 $ & $\beta=4.913$ & $\beta=5.378$ & $\beta=5.450$ & $\beta=5.653$ & $\beta=5.998$  \\
\hline\hline
$t/a=2$  & $0.3752_{-19}^{+19}$ &  $0.4355_{-18}^{+17}$ &  $0.5089_{-18}^{+17}$ & $0.5161_{-13}^{+13}$ & $0.5403_{-13}^{+13}$ & $0.5784_{-11}^{+12}$  \\     
$t/a=4$  & $0.2111_{-13}^{+13}$ &  $0.2678_{-13}^{+14}$ &  $0.3450_{-15}^{+15}$ & $0.3530_{-11}^{+11}$ & $0.3805_{-12}^{+13}$ & $0.4235_{-11}^{+11}$  \\     
$t/a=6$  & $0.1626_{-11}^{+11}$ &  $0.2194_{-12}^{+12}$ &  $0.2983_{-15}^{+15}$ & $0.3071_{-11}^{+11}$ & $0.3366_{-12}^{+12}$ & $0.3814_{-12}^{+11}$  \\     
$t/a=8$  & $0.1417_{-10}^{+10}$ &  $0.1991_{-12}^{+13}$ &  $0.2798_{-15}^{+15}$ & $0.2884_{-12}^{+12}$ & $0.3190_{-13}^{+13}$ & $0.3640_{-12}^{+12}$  \\     
$t/a=10$ & $0.1325_{-09}^{+09}$ &  $0.1899_{-11}^{+12}$ &  $0.2700_{-14}^{+15}$ & $0.2798_{-12}^{+12}$ & $0.3098_{-12}^{+12}$ & $0.3549_{-12}^{+11}$  \\     
$t/a=12$ & $0.1415_{-11}^{+10}$ &  $0.1996_{-12}^{+13}$ &  $0.2784_{-15}^{+16}$ & $0.2897_{-12}^{+12}$ & $0.3185_{-12}^{+13}$ & $0.3634_{-13}^{+13}$  \\     
$t/a=14$ & $0.1625_{-12}^{+11}$ &  $0.2207_{-13}^{+13}$ &  $0.2971_{-16}^{+16}$ & $0.3085_{-12}^{+12}$ & $0.3369_{-12}^{+12}$ & $0.3803_{-13}^{+12}$  \\     
$t/a=16$ & $0.2097_{-13}^{+13}$ &  $0.2695_{-14}^{+14}$ &  $0.3434_{-16}^{+16}$ & $0.3543_{-12}^{+13}$ & $0.3810_{-11}^{+12}$ & $0.4224_{-12}^{+12}$  \\     
$t/a=18$ & $0.3721_{-20}^{+18}$ &  $0.4364_{-19}^{+19}$ &  $0.5075_{-18}^{+18}$ & $0.5174_{-13}^{+14}$ & $0.5405_{-12}^{+13}$ & $0.5779_{-12}^{+12}$  \\     
\hline                                  
\end{tabular}
\end{center}
\end{table}

\begin{table}[h]
\begin{center}
\caption{The Data of $Z_P$ factor ($L/a=12,T/a=24$)} 
\begin{tabular}{|c|c|c|c|c|c|c|}
\hline
{} &   $ \beta=4.600 $ & $\beta=4.913$ & $\beta=5.378$ & $\beta=5.588$ & $\beta=5.786$ & $\beta=6.121$  \\
\hline\hline
$t/a=2$  & $0.3707_{-16}^{+16}$ &  $0.4322_{-15}^{+15}$ &  $0.5046_{-14}^{+15}$ & $0.5322_{-16}^{+17}$ & $0.5544_{-14}^{+15}$ & $0.5875_{-14}^{+13}$  \\     
$t/a=4$  & $0.2109_{-10}^{+11}$ &  $0.2683_{-12}^{+12}$ &  $0.3430_{-12}^{+12}$ & $0.3728_{-15}^{+15}$ & $0.3983_{-14}^{+14}$ & $0.4363_{-14}^{+12}$  \\     
$t/a=6$  & $0.1663_{-09}^{+09}$ &  $0.2242_{-11}^{+11}$ &  $0.3013_{-12}^{+12}$ & $0.3324_{-15}^{+16}$ & $0.3591_{-15}^{+15}$ & $0.3991_{-15}^{+13}$  \\     
$t/a=8$  & $0.1483_{-09}^{+09}$ &  $0.2071_{-11}^{+12}$ &  $0.2857_{-13}^{+13}$ & $0.3180_{-16}^{+16}$ & $0.3446_{-16}^{+16}$ & $0.3857_{-16}^{+15}$  \\     
$t/a=10$ & $0.1370_{-09}^{+09}$ &  $0.1960_{-11}^{+12}$ &  $0.2749_{-12}^{+12}$ & $0.3067_{-16}^{+16}$ & $0.3336_{-16}^{+16}$ & $0.3744_{-16}^{+15}$  \\     
$t/a=12$ & $0.1313_{-09}^{+09}$ &  $0.1909_{-10}^{+11}$ &  $0.2686_{-12}^{+12}$ & $0.2993_{-16}^{+16}$ & $0.3281_{-16}^{+15}$ & $0.3678_{-15}^{+15}$  \\     
$t/a=14$ & $0.1360_{-09}^{+09}$ &  $0.1968_{-11}^{+12}$ &  $0.2740_{-14}^{+13}$ & $0.3043_{-16}^{+16}$ & $0.3350_{-16}^{+15}$ & $0.3745_{-15}^{+15}$  \\     
$t/a=16$ & $0.1471_{-10}^{+10}$ &  $0.2081_{-11}^{+12}$ &  $0.2852_{-13}^{+13}$ & $0.3157_{-16}^{+16}$ & $0.3464_{-15}^{+15}$ & $0.3864_{-14}^{+14}$  \\     
$t/a=18$ & $0.1651_{-10}^{+10}$ &  $0.2254_{-12}^{+12}$ &  $0.3011_{-13}^{+13}$ & $0.3318_{-15}^{+15}$ & $0.3610_{-14}^{+14}$ & $0.4006_{-13}^{+14}$  \\     
$t/a=20$ & $0.2095_{-11}^{+11}$ &  $0.2702_{-12}^{+12}$ &  $0.3429_{-12}^{+12}$ & $0.3729_{-14}^{+14}$ & $0.4003_{-13}^{+13}$ & $0.4381_{-12}^{+13}$  \\     
$t/a=22$ & $0.3694_{-16}^{+16}$ &  $0.4338_{-16}^{+15}$ &  $0.5043_{-14}^{+14}$ & $0.5313_{-16}^{+16}$ & $0.5557_{-14}^{+14}$ & $0.5891_{-12}^{+12}$  \\     
\hline                                  
\end{tabular}
\end{center}
\end{table}

\begin{table}[h]
\begin{center}
\caption{The Data of $Z_P$ factor ($L/a=16,T/a=32$)} 
\begin{tabular}{|c|c|c|c|c|c|c|}
\hline
{} &   $ \beta=5.181 $ & $\beta=5.414$ & $\beta=5.709$ & $\beta=5.796$ & $\beta=5.909$ & $\beta=6.241$  \\
\hline\hline
$t/a=2$  & $0.4779_{-15}^{+15}$ &  $0.5106_{-16}^{+16}$ &  $0.5477_{-18}^{+17}$ & $0.5551_{-15}^{+14}$ & $0.5691_{-17}^{+17}$ & $0.5990_{-17}^{+17}$  \\     
$t/a=4$  & $0.3156_{-13}^{+13}$ &  $0.3519_{-14}^{+14}$ &  $0.3915_{-16}^{+16}$ & $0.4002_{-14}^{+13}$ & $0.4181_{-17}^{+16}$ & $0.4523_{-16}^{+16}$  \\     
$t/a=6$  & $0.2762_{-12}^{+13}$ &  $0.3141_{-14}^{+13}$ &  $0.3566_{-17}^{+17}$ & $0.3655_{-13}^{+13}$ & $0.3852_{-19}^{+18}$ & $0.4204_{-18}^{+17}$  \\     
$t/a=8$  & $0.2627_{-13}^{+13}$ &  $0.3013_{-15}^{+14}$ &  $0.3448_{-19}^{+18}$ & $0.3544_{-14}^{+14}$ & $0.3744_{-20}^{+19}$ & $0.4104_{-19}^{+19}$  \\     
$t/a=10$ & $0.2534_{-13}^{+14}$ &  $0.2923_{-16}^{+16}$ &  $0.3357_{-20}^{+19}$ & $0.3461_{-15}^{+15}$ & $0.3653_{-21}^{+20}$ & $0.4021_{-20}^{+20}$  \\     
$t/a=12$ & $0.2454_{-13}^{+14}$ &  $0.2845_{-16}^{+16}$ &  $0.3275_{-21}^{+19}$ & $0.3382_{-16}^{+16}$ & $0.3564_{-22}^{+21}$ & $0.3933_{-21}^{+20}$  \\     
$t/a=14$ & $0.2382_{-13}^{+14}$ &  $0.2770_{-16}^{+16}$ &  $0.3198_{-20}^{+19}$ & $0.3305_{-16}^{+16}$ & $0.3482_{-22}^{+21}$ & $0.3849_{-20}^{+21}$  \\     
$t/a=16$ & $0.2342_{-14}^{+14}$ &  $0.2739_{-15}^{+15}$ &  $0.3160_{-19}^{+18}$ & $0.3268_{-16}^{+16}$ & $0.3439_{-21}^{+20}$ & $0.3817_{-19}^{+20}$  \\     
$t/a=18$ & $0.2372_{-14}^{+14}$ &  $0.2781_{-15}^{+16}$ &  $0.3203_{-19}^{+19}$ & $0.3306_{-17}^{+16}$ & $0.3479_{-22}^{+21}$ & $0.3870_{-20}^{+21}$  \\     
$t/a=20$ & $0.2441_{-15}^{+14}$ &  $0.2858_{-15}^{+15}$ &  $0.3286_{-20}^{+19}$ & $0.3381_{-17}^{+16}$ & $0.3563_{-22}^{+21}$ & $0.3961_{-20}^{+21}$  \\     
$t/a=22$ & $0.2520_{-14}^{+14}$ &  $0.2940_{-14}^{+15}$ &  $0.3374_{-19}^{+19}$ & $0.3461_{-16}^{+16}$ & $0.3648_{-22}^{+20}$ & $0.4049_{-19}^{+20}$  \\     
$t/a=24$ & $0.2614_{-14}^{+13}$ &  $0.3029_{-14}^{+14}$ &  $0.3466_{-19}^{+18}$ & $0.3551_{-15}^{+15}$ & $0.3733_{-21}^{+20}$ & $0.4135_{-19}^{+19}$  \\     
$t/a=26$ & $0.2752_{-13}^{+13}$ &  $0.3158_{-13}^{+13}$ &  $0.3581_{-18}^{+17}$ & $0.3665_{-13}^{+14}$ & $0.3842_{-19}^{+19}$ & $0.4243_{-17}^{+18}$  \\     
$t/a=28$ & $0.3153_{-13}^{+13}$ &  $0.3536_{-12}^{+13}$ &  $0.3937_{-17}^{+17}$ & $0.4012_{-12}^{+13}$ & $0.4173_{-17}^{+17}$ & $0.4564_{-16}^{+17}$  \\     
$t/a=30$ & $0.4757_{-15}^{+15}$ &  $0.5115_{-15}^{+16}$ &  $0.5486_{-18}^{+18}$ & $0.5550_{-13}^{+14}$ & $0.5686_{-17}^{+17}$ & $0.6018_{-16}^{+16}$  \\     
\hline                                  
\end{tabular}
\end{center}
\end{table}

\begin{table}[h]
\begin{center}
\caption{The Data of $Z_P$ factor ($L/a=20,T/a=40$)} \label{table:nonpert-Z-L20}
\begin{tabular}{|c|c|c|c|c|c|c|}
\hline
{} &   $ \beta=5.450$ & $\beta=5.653$ & $\beta=5.733$ & $\beta=5.944$ & $\beta=5.998$ & $\beta=6.296$  \\
\hline\hline
$t/a=2$  & $0.5139 _{-18} ^{+18}$ &  $0.5382 _{-18} ^{+18}$ &  $0.5504 _{-19} ^{+19}$ & $0.5762 _{-19} ^{+18}$ & $0.5733 _{-15} ^{+15}$ & $0.6049_{-18}^{+19}$  \\     
$t/a=4$  & $0.3552 _{-17} ^{+17}$ &  $0.3830 _{-16} ^{+17}$ &  $0.3961 _{-17} ^{+18}$ & $0.4257 _{-18} ^{+18}$ & $0.4239 _{-15} ^{+15}$ & $0.4605_{-19}^{+18}$  \\     
$t/a=6$  & $0.3199 _{-17} ^{+17}$ &  $0.3492 _{-16} ^{+18}$ &  $0.3623 _{-18} ^{+19}$ & $0.3933 _{-20} ^{+19}$ & $0.3919 _{-16} ^{+16}$ & $0.4313_{-20}^{+19}$  \\     
$t/a=8$  & $0.3089 _{-18} ^{+18}$ &  $0.3389 _{-18} ^{+19}$ &  $0.3522 _{-19} ^{+20}$ & $0.3842 _{-21} ^{+20}$ & $0.3821 _{-17} ^{+17}$ & $0.4230_{-21}^{+21}$  \\     
$t/a=10$ & $0.3008 _{-19} ^{+18}$ &  $0.3312 _{-19} ^{+20}$ &  $0.3448 _{-20} ^{+21}$ & $0.3765 _{-22} ^{+22}$ & $0.3741 _{-17} ^{+17}$ & $0.4162_{-22}^{+22}$  \\     
$t/a=12$ & $0.2932 _{-19} ^{+19}$ &  $0.3240 _{-20} ^{+20}$ &  $0.3378 _{-21} ^{+21}$ & $0.3683 _{-24} ^{+23}$ & $0.3665 _{-18} ^{+18}$ & $0.4093_{-22}^{+22}$  \\     
$t/a=14$ & $0.2862 _{-20} ^{+19}$ &  $0.3170 _{-20} ^{+21}$ &  $0.3306 _{-22} ^{+22}$ & $0.3605 _{-25} ^{+25}$ & $0.3585 _{-18} ^{+19}$ & $0.4026_{-22}^{+23}$  \\     
$t/a=16$ & $0.2796 _{-19} ^{+19}$ &  $0.3103 _{-20} ^{+21}$ &  $0.3231 _{-22} ^{+22}$ & $0.3527 _{-25} ^{+25}$ & $0.3510 _{-18} ^{+19}$ & $0.3953_{-22}^{+23}$  \\     
$t/a=18$ & $0.2739 _{-18} ^{+19}$ &  $0.3047 _{-20} ^{+20}$ &  $0.3164 _{-22} ^{+22}$ & $0.3460 _{-24} ^{+26}$ & $0.3452 _{-18} ^{+19}$ & $0.3888_{-22}^{+23}$  \\     
$t/a=20$ & $0.2713 _{-17} ^{+19}$ &  $0.3024 _{-19} ^{+19}$ &  $0.3130 _{-22} ^{+22}$ & $0.3432 _{-25} ^{+25}$ & $0.3437 _{-18} ^{+19}$ & $0.3852_{-22}^{+23}$  \\     
$t/a=22$ & $0.2735 _{-19} ^{+19}$ &  $0.3057 _{-19} ^{+20}$ &  $0.3151 _{-23} ^{+23}$ & $0.3463 _{-25} ^{+26}$ & $0.3481 _{-18} ^{+19}$ & $0.3877_{-23}^{+24}$  \\     
$t/a=24$ & $0.2791 _{-20} ^{+19}$ &  $0.3125 _{-19} ^{+20}$ &  $0.3209 _{-23} ^{+24}$ & $0.3527 _{-25} ^{+26}$ & $0.3558 _{-18} ^{+19}$ & $0.3941_{-24}^{+24}$  \\     
$t/a=26$ & $0.2858 _{-20} ^{+19}$ &  $0.3194 _{-19} ^{+20}$ &  $0.3282 _{-23} ^{+24}$ & $0.3598 _{-24} ^{+25}$ & $0.3637 _{-18} ^{+19}$ & $0.4015_{-23}^{+23}$  \\     
$t/a=28$ & $0.2931 _{-19} ^{+19}$ &  $0.3265 _{-19} ^{+19}$ &  $0.3356 _{-22} ^{+23}$ & $0.3673 _{-23} ^{+24}$ & $0.3713 _{-18} ^{+19}$ & $0.4090_{-23}^{+22}$  \\     
$t/a=30$ & $0.3007 _{-18} ^{+18}$ &  $0.3337 _{-18} ^{+19}$ &  $0.3436 _{-21} ^{+22}$ & $0.3746 _{-22} ^{+23}$ & $0.3788 _{-18} ^{+18}$ & $0.4159_{-22}^{+21}$  \\     
$t/a=32$ & $0.3090 _{-17} ^{+17}$ &  $0.3409 _{-17} ^{+18}$ &  $0.3517 _{-20} ^{+21}$ & $0.3816 _{-22} ^{+23}$ & $0.3865 _{-18} ^{+17}$ & $0.4219_{-21}^{+20}$  \\     
$t/a=34$ & $0.3204 _{-16} ^{+16}$ &  $0.3502 _{-16} ^{+17}$ &  $0.3625 _{-19} ^{+19}$ & $0.3902 _{-21} ^{+21}$ & $0.3952 _{-17} ^{+16}$ & $0.4289_{-21}^{+19}$  \\     
$t/a=36$ & $0.3566 _{-15} ^{+15}$ &  $0.3840 _{-15} ^{+16}$ &  $0.3976 _{-18} ^{+18}$ & $0.4213 _{-21} ^{+21}$ & $0.4266 _{-16} ^{+15}$ & $0.4574_{-20}^{+19}$  \\     
$t/a=38$ & $0.5156 _{-18} ^{+18}$ &  $0.5396 _{-18} ^{+17}$ &  $0.5532 _{-20} ^{+20}$ & $0.5702 _{-21} ^{+21}$ & $0.5754 _{-17} ^{+16}$ & $0.6030_{-19}^{+18}$  \\     
\hline                                  
\end{tabular}
\end{center}
\end{table}

\newpage

\end{document}